\documentclass[nofootinbib,preprint,tightenlines,superscriptaddress]{revtex4}%

\usepackage{amsfonts}
\usepackage[utf8]{inputenc}
\usepackage{amssymb,amsthm,amsmath,amstext,amsbsy,bbm}
\usepackage{url}
\usepackage{nicefrac}
\usepackage{graphicx}
\usepackage{ifpdf}
\usepackage[multiple]{footmisc}%
\usepackage{amsmath}%
\usepackage{amssymb}
\usepackage{color}

\newcommand{\ie}{\textit{i.e.}}

\newcommand{\cf}{\textit{cf.}}

\newcommand{\etal}{\textit{et al.}}

\newcommand{\mathspace}{\ \ }
\newcommand{\mathtext}[1]{\mathspace\text{#1}\mathspace}

\newcommand{\dd}{\mathrm{d}}
\newcommand{\ii}{\mathrm{i}}
\newcommand{\ee}{\mathrm{e}}

\newcommand{\OO}{\mathcal{O}}

\newcommand{\Rp}{\mathrm{Re}}

\newcommand{\keV}{\ensuremath{\mathrm{keV}}}
\newcommand{\MeV}{\ensuremath{\mathrm{MeV}}}
\newcommand{\fm}{\ensuremath{\mathrm{fm}}}

\begin{document}

\title{Causality constraints for charged particles}

\author{Sebastian K\"onig}
\email{koenig@hiskp.uni-bonn.de}
\affiliation{Helmholtz-Institut f\"ur Strahlen- und Kernphysik (Theorie)\\
and Bethe Center for Theoretical Physics, Universit\"at Bonn, 53115 Bonn,
Germany\\[0.5em]}

\author{Dean Lee}
\email{dean_lee@ncsu.edu}
\affiliation{Department of Physics, North Carolina State University, Raleigh,
NC 27695, USA}
\affiliation{Helmholtz-Institut f\"ur Strahlen- und Kernphysik (Theorie)\\
and Bethe Center for Theoretical Physics, Universit\"at Bonn, 53115 Bonn,
Germany\\[0.5em]}

\author{H.-W. Hammer}
\email{hammer@hiskp.uni-bonn.de}
\affiliation{Helmholtz-Institut f\"ur Strahlen- und Kernphysik (Theorie)\\
and Bethe Center for Theoretical Physics, Universit\"at Bonn, 53115 Bonn,
Germany\\[0.5em]}

\date{\today}

\begin{abstract}
In quantum systems with short-range interactions, causality imposes nontrivial
constraints on low-energy scattering parameters.  We investigate these causality
constraints for systems where a long-range Coulomb potential is present in
addition to a short-range interaction.  The main result is an upper bound for
the Coulomb-modified effective range parameter.  We discuss the implications of
this bound to the effective field theory for nuclear halo systems.  In
particular, we consider several examples of proton--nucleus and nucleus--nucleus
scattering.  For the bound-state regime, we find relations for the asymptotic
normalization coefficients (ANCs) of nuclear halo states. As an application of
these relations, we extract the ANCs of the excited $2^+$ and $1^-$ states in
$^{16}$O from $\alpha-^{12}$C scattering data.
\end{abstract}

\maketitle

\section{Introduction}
\label{sec:Intro}

The constraints of causality for two-body scattering with finite-range
interactions were first derived by Wigner~\cite{Wigner:1955zz}.  The causality
bound can be understood as a lower bound on the time delay between
the incoming and outgoing wave packets, $\Delta t$.  When $\Delta t$ is
negative, the outgoing wave packet departs earlier than for the non-interacting
system.  However, the incoming wave must first reach the interaction region
before the outgoing wave can leave.  In low-energy scattering this manifests
itself as an upper bound on the effective range parameter.  In
Ref.~\cite{Phillips:1996ae}, Phillips and Cohen derived this bound for S-wave 
scattering with finite-range interactions.  Some constraints on
nucleon--nucleon scattering and the chiral two-pion exchange potential were
considered in Ref.~\cite{PavonValderrama:2005wv}, and relations between the
scattering length and effective range have been explored for one-boson
exchange potentials~\cite{Cordon:2009pj} and van der Waals
potentials~\cite{Cordon:2009wh}.  In Refs.~\cite{Hammer:2009zh,Hammer:2010fw}
the causality bounds for finite-range interactions were extended to an
arbitrary number of spacetime dimensions and arbitrary angular momentum.  The
extension to systems with partial wave mixing was first studied in
Ref.~\cite{Elhatisari:2012ym}.

In this paper, we consider the causality constraints for the scattering of two
charged particles with an arbitrary finite-range interaction.  Our analysis
is the first study of causality bounds that takes into account the long-range
Coulomb force.  The results presented here are relevant to studies of
low-energy scattering of nuclei and nucleons using effective field theory (EFT).
In particular, they are important for the application of effective field theory
to nuclear halo systems (cf.~Refs.~\cite{Zhukov:1993aw,Riisager:1994zz}).

This so-called halo EFT utilizes the separation of energy scales between the
internal excitations of the core nucleus and soft halo physics.  An effective
Lagrangian is built order by order from local interactions and used to
describe low-energy phenomena such as shallow bound states, charge radii,
near-threshold resonances, radiative capture reactions, and soft
photodissociation.  Halo EFT has been used to describe neutron--alpha
scattering~\cite{Bertulani:2002sz,Bedaque:2003wa} and alpha--alpha
scattering~\cite{Higa:2008dn}.  Bound single-neutron halo systems such as 
$^{11}$Be~\cite{Hammer:2011ye} and $^{8}$Li~\cite{Rupak:2011nk,Fernando:2011ts}
have been studied as well as various two-neutron halo
systems~\cite{Canham:2008jd,Canham:2009xg,Rotureau:2012yu}.

There is an important connection between causality bounds and the convergence
of effective field theory calculations with increasing
order~\cite{Elhatisari:2012ym}.  For local contact interactions, the range of
the effective interaction is controlled by the momentum cutoff scale of the
effective theory.  In effective theories with non-perturbative renormalization,
which typically occur in nuclear physics, exact cutoff-independence can
generally not be achieved.  There is a ``natural'' value of the cutoff at which
all higher-order corrections scale as expected from dimensional analysis.  If
the cutoff is taken larger, ``new physics'' intervenes, the corrections scale
unnaturally, and unitarity violations may occur.  This is different from what
one encounters in high-energy particle physics where the renormalization is
typically perturbative and cutoff momenta can be taken arbitrarily large.  For
calculations using dimensional regularization, the renormalization scale plays a
similar role in regulating ultraviolet physics.

The term ``new physics'', in this context, refers to details left out
(integrated out) in the effective theory.  In the case of halo EFT, these
details are the finite size of the core nucleus and its internal excitations
as well as the exponential tail of the pion-exchange interaction.  Problems with
convergence of the effective theory can occur if the cutoff scale is set higher
than the scale of the new physics. However, it is useful to have a more
quantitative measure of when problems may appear, and this is where the
causality bound provides a useful diagnostic tool.  For each scattering channel
we use the physical scattering parameters to compute a quantity called the
causal range, $R_{c}$.  It is the minimum range for finite-range interactions
consistent with the requirements of causality and unitarity.  For any fixed
cutoff scale, the causality bound marks a branch cut of the effective theory
when viewed as a function of physical scattering
parameters~\cite{Elhatisari:2012ym}.  The coupling constants of the effective
theory become complex when scattering parameters violating the causality bound
are enforced.  These branch cuts do not appear in perturbation theory; however,
a nearby branch point can spoil the absolute convergence of the perturbative
expansion.  

Our results can be viewed as a guide for improving the convergence of halo EFT
calculations.  In particular, if the cutoff momentum used in a calculation is
too high, then problems with convergence may appear in some observables.  
Consequently, the causal range can be used to estimate the  ``natural'' 
ultraviolet cutoff $\Lambda$ of the effective theory as  $R_{c}^{-1}$.  The
natural cutoff is optimal in the sense that no known infrared physics is left
out of the theory and that all corrections involving the ultraviolet cutoff
scale naturally~\cite{Lepage:1989hf,Lepage:1997cs}.  Increasing the cutoff
beyond the natural value will not improve the accuracy of the calculation. 

The causality bounds also have an impact in the regime of bound states.  For
two-body halo states, or more generally whenever there is a shallow two-body
bound state close to threshold, the same integral identity that yields the
causality bound for the effective range can be used to derive a relation
between the asymptotic normalization constant (ANC) of the bound-state wave
function, the binding momentum, and the effective range for the scattering of
the two halo constituents.  This relation can be shown to be equivalent to a
result previously derived by Sparenberg~\etal~\cite{Sparenberg:2009rv}.  Its
significance lies in the fact that the ANC is an important input parameter for
the calculation of near-threshold radiative capture and photodissociation
reactions.  The causality bounds also constrain the range of model potentials
that are fitted to scattering data in order to extract ANCs.

\medskip
The organization of this paper is as follows.  We first review the general
theory of scattering for two charged particles with additional short-range
interactions.  Our analysis includes both attractive and repulsive Coulomb
forces.  In the next section we derive the charged-particle causality bounds
for arbitrary values of the orbital angular momentum.  Using the causality
bounds, we extract and discuss the causal range for several nuclear scattering
processes including proton--proton, proton--deuteron, proton--$^{3}$He,
proton--alpha, and alpha--alpha scattering.  We then elucidate the relation for
the ANC and extract the ANCs of the excited $2^+$ and $1^-$ states in $^{16}$O
from $\alpha-^{12}$C scattering data as an application.  We conclude with a
summary of the main results and provide an outlook.  In the appendices, we
provide technical details of the derivation as well as numerical examples.

\section{Two charged particles with short-range interactions}
\label{sec:Preliminaries}

In this section, we review some preliminaries that we will need later in order
to derive the desired causality bounds.  We consider a two-particle system with
reduced mass $\mu$ interacting via a finite-range potential with range $R$.
We write the interaction as a real symmetric operator with kernel
$V(r,r^{\prime})$ satisfying the finite-range condition,
\begin{equation}
 V(r,r') = 0 \mathtext{if} r>R \mathtext{or} r'>R \,.
\label{eq:V-FR}
\end{equation}
In particular, we assume that the interaction is energy-independent. 
After giving a detailed formal derivation of the causality bounds in what
follows, we will come back to the question what the above assumptions mean for
the application to (halo) EFT calculations in Sec.~\ref{sec:Results}.

\medskip
In the absence of Coulomb interactions the system is described by the radial
Schr\"{o}dinger equation,
\begin{equation}
 p^2 u_\ell(r) = -\frac{\dd^2}{\dd r^2} u_\ell(r)
 + \frac{\ell(\ell+1)}{r^2} u_\ell(r)
 + 2\mu\int_0^R\dd r'\,V(r,r')\,u_\ell(r') \,,
\label{eq:SG-rad-u}
\end{equation}
where by $u_\ell^{(p)}$ we explicitly indicate a solution for center-of-mass
momentum $p$ in the following.  Following the conventions of
Ref.~\cite{Hammer:2010fw}, the normalization of $u_{\ell}^{(p)}$ is chosen such
that for $r\geq R$ we have
\begin{equation}
 u_\ell^{(p)}(r) = p^\ell
 \left[\cot\delta_\ell(p)\,S_\ell(pr)+C_\ell(pr)\right] \,,
\label{eq:u-asympt}
\end{equation}
where $S_{\ell}$ and $C_{\ell}$ are the Riccati-Bessel functions and
$\delta_\ell(p)$ is the scattering phase shift.

\medskip
If the particles carry electromagnetic charges $Z_{1}e$ and $Z_{2}e$,
respectively, there is a Coulomb potential in addition to the finite-range
interaction. The radial Schr\"{o}dinger equation now reads
\begin{equation}
 p^2 w_\ell(r) = -\frac{\dd^2}{\dd r^2} w_\ell(r)
 + \frac{\ell(\ell+1)}{r^2} w_\ell(r)
 + 2\mu\int_0^R\dd r'\,V(r,r')\,w_\ell(r')
 + \frac{\gamma}{r} w_\ell(r)
\label{eq:SG-rad-w}
\end{equation}
with the Coulomb parameter
\begin{equation}
 \gamma = 2\mu \cdot\alpha Z_1 Z_2 \,.
\end{equation}
The normalization of $w_{\ell}^{(p)}$ is now chosen such that for
$r\geq R$ we have
\begin{equation}
 w_\ell^{(p)}(r) = p^\ell C_{\eta,\ell}
 \left[\cot\tilde{\delta}_\ell(p)\,F_\ell^{(p)}(r)+G_\ell^{(p)}(r)\right] \,,
\label{eq:w-asympt}
\end{equation}
where $F_{\ell}^{(p)}$ and $G_{\ell}^{(p)}$ are the regular and irregular
Coulomb wave functions~\cite{Yost:1936zz}, respectively, and
$\tilde{\delta}_{\ell}$ is the phase shift of the full solution $w_{\ell}^{(p)}$
compared to the regular Coulomb function
$F_{\ell}^{(p)}$~\cite{Bethe:1949yr}.\footnote{Note that for $\ell=0$ our
normalization is the same as chosen in Ref.~\cite{Bethe:1949yr}, \textit{i.e.},
for $r\geq R$ our solution $w_{0}^{(p)}$ coincides with the function $\varphi$
defined in Eq.~(42) of that paper.}  The factor $C_{\eta,\ell}$ is given by
\begin{equation}
 \eta = \frac{\gamma}{2p} \,,
\label{eq:eta}
\end{equation}
\begin{equation}
 C_{\eta,0}^2=\frac{2\pi\eta}{\ee^{2\pi\eta}-1} \,,
\label{eq:C_eta_zero}
\end{equation}
and
\begin{equation}
 C_{\eta,\ell}^2 = \frac{2^{2\ell}}{\left[(2\ell+1)!\right]^2}
 \prod\limits_{s=1}^\ell (s^2+\eta^2) \cdot C_{\eta,0}^2 \,.
\label{eq:C_eta}
\end{equation}

\subsection{Coulomb wave functions}
\label{sec:CoulombWF}

For the Coulomb wave functions $F_{\ell}^{(p)}$ and $G_{\ell}^{(p)}$ we use the
conventions introduced by Yost, Breit and Wheeler in Ref.~\cite{Yost:1936zz}. 
Explicitly, we write them as~\cite{Slater,Boersma:1969ab}
\begin{subequations}%
\begin{equation}
 F_\ell^{(p)}(r) = \frac12\left|\frac{\ee^{\ii\frac\pi2\kappa}\,
 \Gamma\!\left(\frac12+m-\kappa\right)}{\Gamma(2m+1)}\right|
 \ee^{-\ii\frac\pi2(\frac12+m)} M_{\kappa,m}(z)
\label{eq:F}
\end{equation}
\begin{equation}
 G_\ell^{(p)}(r) = \frac{\Gamma\!\left(\frac12+m-\kappa\right)}
 {\left|\Gamma\!\left(\frac12+m-\kappa\right)\right|}
 \ee^{-\ii\frac\pi2(\frac12-m+\kappa)} W_{\kappa,m}(z)
 + \ii F_\ell^{(p)}(r) \,,
\label{eq:G}
\end{equation}
\label{eq:FG}%
\end{subequations}%
where
\begin{equation}
 \rho = p \cdot r \mathtext{,} z = 2\ii\rho \mathtext{,}
 \kappa=\ii\eta \mathtext{,} m = \ell+\frac12 \,.
\label{eq:rho-z-kappa-m}
\end{equation}
The functions $M_{\kappa,m}$ and $W_{\kappa,m}$ are Whittaker functions, which
can be expressed in terms of hypergeometric functions as
\begin{equation}
 M_{\kappa,m}(z) = \ee^{-\frac12z}z^{\frac12+m}
 {_1F_1}\!\left(\tfrac12+m-\kappa,1+2m;z\right) \,,
\label{eq:WhittakerM}
\end{equation}%
\begin{equation}
 W_{\kappa,m}(z) = \ee^{-\frac12z}z^{\frac12+m}
 U\!\left(\tfrac12+m-\kappa,1+2m;z\right) \,.
\label{eq:WhittakerW}
\end{equation}
$_{1}F_{1}(a,b;z)$ is Kummer's function of the first kind,
\begin{equation}
 _1F_1(a,b;z) = \sum_{n=0}^\infty\frac{a^{(n)}z^n}{b^{(n)}n!}
 \mathtext{,} a^{(n)} = a(a+1)\cdots(a+n-1) \,,
\end{equation}
and
\begin{equation}
 U(a,b;z) = \frac{\Gamma(1-b)}{\Gamma(1+a-b)}{_1F_1}(a,b;z)
 + \frac{\Gamma(b-1)}{\Gamma(a)}z^{1-b}{_1F_1}(a-b+1,2-b;z) \,.
\end{equation}
A comprehensive discussion of the functions $F_{\ell}^{(p)}(r)$ and
$G_{\ell}^{(p)}(r)$ can be found in Ref.~\cite{Hull:1959ab}.  For asymptotically
large $\rho=pr,$ the Coulomb wave functions behave as
\begin{align}
 F_\ell^{(p)}(r) &\sim \sin(\rho-\ell\pi/2-\eta\log(2\rho)+\sigma_\ell) \,, \\
 G_\ell^{(p)}(r) &\sim \cos(\rho-\ell\pi/2-\eta\log(2\rho)+\sigma_\ell)
\label{eq:FG-rho-infty}
\end{align}
with the Coulomb phase shift
\begin{equation}
 \sigma_\ell = \arg\Gamma(\ell+1+\ii\eta) \,.
\label{eq:sigma-ell-simple}
\end{equation}
In the limit $\rho\rightarrow0$, their behavior is
\begin{subequations}
\label{eq:FG-rho-zero}
\begin{align}
 \label{eq:F-rho-zero}
 F_\ell^{(p)}(r) &\sim C_{\eta,\ell}\,\rho^{\ell+1} \,, \\
 \label{eq:G-rho-zero}
 G_\ell^{(p)}(r) &\sim \frac{\rho^{-\ell}}{C_{\eta,\ell}(2\ell+1)} \,.
\end{align}
\end{subequations}

\subsection{Effective range expansions}
\label{sec:ERE}

For a system without Coulomb force and only a short-range interaction, one has
the well-known effective range expansion
\begin{equation}
 p^{2\ell+1}\cot\delta_\ell(p)=-\frac1{a_\ell}+\frac12r_\ell\,p^2 + \cdots \,,
\label{eq:ERE}
\end{equation}
where $a_{\ell}$ and $r_{\ell}$ are the scattering and effective range
parameters, respectively.  For a system of charged particles we have instead the
Coulomb-modified effective range expansion
\begin{equation}
 C_{\eta,\ell}^2\,p^{2\ell+1}\,\cot\tilde{\delta}_\ell(p)
 + \gamma\,h_{\ell}(p)
 = -\frac1{a^C_\ell}+\frac12r^C_\ell\,p^2 + \cdots \,,
\label{eq:ERE-CbMod}
\end{equation}
where
\begin{equation}
 h_{\ell}(p) = p^{2\ell}\frac{C_{\eta,\ell}^2}{C_{\eta,0}^2}\,h(\eta) \,,
\label{eq:h-ell}
\end{equation}
\begin{equation}
 h(\eta) = \mathrm{Re}\,\psi(\ii\eta)-\log|\eta| \,,
\label{eq:h}
\end{equation}
and $\psi(z)={\Gamma^{\prime}(z)}/{\Gamma(z)}$ is the logarithmic derivative
of the Gamma function, also called digamma function.  For $\ell=0$, the
expansion simplifies to
\begin{equation}
 C_{\eta,0}^2\,p\,\cot\tilde{\delta}_0(p) + \gamma\,h(\eta)
 = -\frac1{a_0^C}+\frac12r_0^C\,p^2 + \cdots \,.
\label{eq:ERE-CbMod-0}
\end{equation}
A derivation of Eq.~\eqref{eq:ERE-CbMod-0} for the case of proton--proton
scattering can be found, for example, in Ref.~\cite{Bethe:1949yr}.  See also
Ref.~\cite{Jackson:1950zz} for a detailed discussion.  The analytic properties
of the $\ell=0$ modified effective range function are investigated in
Ref.~\cite{vanHaeringen:1981pb}.\footnote{As a remark we note that on first
sight the expansion given in Eq.~(51) of Bethe's paper~\cite{Bethe:1949yr}
seems to be different from the one given here in Eq.~\eqref{eq:ERE-CbMod-0},
which is the same as given in later publications referring to Bethe's result.
The $\eta$-dependent function on the left hand side of Bethe's expansion
appears to differ from our $h(\eta)$ by two times the Euler-Mascheroni
constant $\gamma_{E}$.  This apparent conflict can be resolved by noting that
the $g(\eta)$ in Eq.~(51) of Ref.~\cite{Bethe:1949yr} is \emph{not} the
function defined in Eq.~(47a) of the same paper, but rather given by
$\lim_{\eta_{1}\rightarrow\infty}[g(\eta)-g(\eta_{1})]$, where in this latter
expression the $g$ from Eq.~(47a) is meant.  The limiting process then yields
exactly the term $-2\gamma_{E}$.}

In Ref.~\cite{Bolle:1984ab}, Boll\'{e} and Gesztesy derived a very general
form of the Coulomb-modified effective range expansion for an arbitrary number
of spatial dimensions.  Specializing their result to the three-dimensional
case, a version of Eq.~\eqref{eq:ERE-CbMod} can be written as
\begin{equation}
 C_{\eta,\ell}^2\,p^{2\ell+1}\left(\cot\tilde{\delta}_\ell(p)-\ii\right)
 + \gamma\,\tilde{h}_\ell(p)
 = -\frac1{a^C_\ell}+\frac12r^C_\ell\,p^2 + \cdots
\label{eq:ERE-CbMod-complex}
\end{equation}
with\footnote{This definition essentially comes from combining Eqs.~(4.1)
and~(4.2) in Ref.~\cite{Bolle:1984ab}, with the correction that the exponent
in Eq.~(4.2) should be $-2$ rather than $2$.}
\begin{equation}
 \tilde{h}_\ell(p) = \frac{(2p)^{2\ell}}{\Gamma(2\ell+2)^2}
 \frac{|\Gamma(\ell+1+\ii\eta)|^2}{|\Gamma(1+\ii\eta)|^2}
 \left(\psi(\ii\eta)+\frac1{2\ii\eta}-\log(\ii\eta)\right) \,.
\label{eq:h-tilde}
\end{equation}
The latter function can be rewritten using
\begin{equation}
 C_{\eta,\ell}^2 = \frac{2^{2\ell}}{\Gamma(2\ell+2)^2}
 \frac{|\Gamma(\ell+1+\ii\eta)|^2}{|\Gamma(1+\ii\eta)|^2} \cdot C_{\eta,0}^2 \,,
\label{eq:C_eta-complex}
\end{equation}
with $C_{\eta,0}^{2}$ as defined in Eq.~\eqref{eq:C_eta_zero}.  The expressions
given here reproduce Eqs.~\eqref{eq:ERE-CbMod} and~\eqref{eq:C_eta} when one
explicitly assumes that the momentum $p$ is real.  In fact, one has to rewrite
Eq.~\eqref{eq:h-tilde} in this manner in order to get an effective range
function that is analytic in $p^2$ around threshold.

\medskip
The conventions for the Coulomb-modified efffective range expansion for general
$\ell$ that we use throughout this work, Eqs.~\eqref{eq:ERE-CbMod}
and~\eqref{eq:ERE-CbMod-complex}, are the same as in Ref.~\cite{Bolle:1984ab}. 
We note, however, that a different convention is also used in the literature
which differs from ours by an overall momentum-independent factor.  The
effective range expansion given in
Refs.~\cite{Hamilton:1973xd,Haeringen:1977ab,deMaag:1984ab} uses the
normalization
\begin{equation}
 \left(\frac{\Gamma(2\ell+2)}{2^{\ell}\,\Gamma(\ell+1)}\right)^{\!2}
 \left[C_{\eta,\ell}^{2}\,p^{2\ell+1}\,
 \left(\cot\tilde{\delta}_{\ell}(p)-\ii\right)
 + \gamma\,\tilde{h}_{\ell}(p)\right] 
 = -\frac{1}{\tilde{a}_{\ell}^{C}}
 + \frac{1}{2}\tilde{r}_{\ell}^{C}\,p^{2}+\cdots \,.
\label{eq:ERE-alternative}
\end{equation}
This choice of normalization has the advantage of having a more direct
correspondence to the ordinary effective range expansion without Coulomb
effects.  For $\ell=0$, both conventions give the same expression.

\section{Causality bounds for charged particles}
\label{sec:CausalityBounds}

With the Coulomb wave functions and Coulomb-modified effective range
expansions at our hands, we can now closely follow the derivation presented in
Ref.~\cite{Hammer:2010fw} for scattering in the absence of Coulomb interactions.

\subsection{Wronskian identities}
\label{sec:WronskianIdentities}

We consider solutions of the radial Schr\"{o}dinger
equation~\eqref{eq:SG-rad-w} for two different momenta $p_{A}$ and $p_{B}$.
Introducing the short-hand notation
\begin{equation}
 w_{A,B}(r) = w^{(p_{A,B})}_\ell(r) \,,
\end{equation}
\textit{i.e.}, suppressing the angular-momentum subscript $\ell$ for
convenience, we get
\begin{multline}
 (p_B^2-p_A^2) \int_\epsilon^r\dd r' w_A(r')w_B(r')
 = (w_B w_A' - w_A w_B')\big|_\epsilon^r \\
 - 2\mu \int_\epsilon^r\dd r'\int_0^R\dd r
 \left[w_B(r)V(r,r')w_A(r') - w_A(r)V(r,r')w_B(r')\right]
 \label{eq:W_BA-w-int-pre}
\end{multline}
by subtracting $w_{A}$ times the equation for $w_{B}$ from that for $w_{B}$
multiplied by $w_{A}$, as it is done in Ref.~\cite{Hammer:2010fw}, and
integrating from a small radius $\epsilon$ to $r$.

We assume that our interaction $V(r,r^{\prime})$ is such that it alone permits
a solution that is sufficiently regular at the origin, \textit{i.e.},
$u_{\ell}(0)=0$ and $\partial_{r}u_{\ell}$ stays finite as $r\rightarrow0$,
where $u_{\ell}$ is a solution of Eq.~\eqref{eq:SG-rad-u}.  As boundary
condition for the solutions $w_{A,B}$ of the full radial Schr\"{o}dinger
equation we can then demand as well that they vanish with finite derivative at
the origin.  If we only had the Coulomb potential and no additional interaction,
this is fulfilled by the regular Coulomb function $F_{\ell}^{(p)}(r)$,
\cf~Eq.~\eqref{eq:F-rho-zero}.  We can thus take the limit $\epsilon\to0$ in
Eq.~\eqref{eq:W_BA-w-int-pre} and get the relation
\begin{equation}
 W[w_B,w_A](r) = (p_B^2-p_A^2)\int_0^r\dd r'\,w_A(r')w_B(r') \,,
 \label{eq:W_BA-w-int}
\end{equation}
where the Wronskian $W[w_B,w_A]$ is defined as
\begin{equation}
 W[w_B,w_A](r) = w_B(r)w_A'(r) - w_A(r)w_B'(r) \,.
\end{equation}

\subsection{Rewriting the wave functions}
\label{sec:RewriteWF}

Following further the derivation presented in Ref.~\cite{Hammer:2010fw}, we
re-express the solutions $w_{\ell}^{(p)}(r)$ in terms of functions $f(p,r)$ and
$g(p,r)$ such that
\begin{equation}
 w_\ell^{(p)}(r) = p^{2\ell+1} C_{\eta,\ell}^2
 \cot\tilde{\delta}_\ell(p)\,f(p,r)+g(p,r)
\label{eq:w-asympt-fg}
\end{equation}
for $r\geq R$, with $f(p,r)$ analytic in $p^{2}$,
\begin{equation}
 f(p,r) = f_0(r) + f_2(r)\,p^2 + \OO(p^4) \,,
\label{eq:f-exp}
\end{equation}
and
\begin{subequations}
\label{eq:g-exp}%
\begin{equation}
  g(p,r) = \tilde{g}(p,r) + \phi(p) \cdot f(p,r) \,.
\end{equation}
The $g(p,r)$ contains a term which is non-analytic in $p^{2}$ and is
proportional to $f(p,r)$.  The remainder $\tilde{g}(p,r)$, however, is analytic
in $p^{2}$,
\begin{equation}
 \tilde{g}(p,r) = g_0(r) + g_2(r)\,p^2 + \OO(p^4) \,.
\end{equation}
\end{subequations}
Explicit expressions for these functions can be obtained from the results of
Boll\'{e} and Gesztesy~\cite{Bolle:1984ab}; they are discussed further in
Appendix~\ref{sec:CoulombWF-BG}.  Combining Eqs.~\eqref{eq:w-asympt}
and~\eqref{eq:w-asympt-fg}, we find
\begin{subequations}%
\begin{equation}
 f(p,r) = \frac{1}{p^{\ell+1}C_{\eta,\ell}} F_\ell^{(p)}(r)
\label{eq:f-F}
\end{equation}
and
\begin{equation}
 g(p,r) = p^{\ell}C_{\eta,\ell}\,G_\ell^{(p)}(r) \,.
\label{eq:g-G}
\end{equation}
\label{eq:fg-FG}%
\end{subequations}%

\medskip
We insert now the modified effective range expansion~\eqref{eq:ERE-CbMod} into
the asymptotic Coulomb wave function~\eqref{eq:w-asympt-fg}.  Note also that
\begin{equation}
 \phi(p) = \gamma\,\tilde{h}_\ell(p) - \ii p^{2\ell+1} C_{\eta,\ell}^2
 = \gamma\,h_\ell(p) \,,
\end{equation}
where the last step follows from the relations in
Appendix~\ref{sec:CoulombWF-BG}.  The term involving $h_{\ell}(p)$ exactly
drops out and we are left with
\begin{equation}
 w_\ell^{(p)}(r) = \left(-\frac1{a^C_\ell}+\frac12r^C_\ell\,p^2 + \cdots\right)
 f(p,r) + \tilde{g}(p,r) \mathtext{,} r \geq R \,.
\end{equation}
Thus it is possible to choose a normalization such that $w_{\ell}^{(p)}(r)$ is
analytic in $p^2$.  Combining this with the expansions~\eqref{eq:f-exp}
and~\eqref{eq:g-exp}, we arrive at
\begin{equation}
 w_\ell^{(p)}(r) = -\frac1{a_\ell^C}f_0(r) + g_0(r)
 + p^2\left[\frac12r_\ell^C f_0(r)-\frac1{a_\ell^C} f_2(r) + g_2(r)\right]
 + \OO(p^4)\,.
\label{eq:w-asympt-ERE}
\end{equation}

\subsection{Causality bounds}

From here, we can proceed exactly as in Ref.~\cite{Hammer:2010fw}.  For the
Wronskian of two solutions $w_{A}$ and $w_{B}$ for $r\geq R$ we find
\begin{multline}
 W[w_B,w_A](r) = (p_B^2-p_A^2)\Bigg\{\frac12r_\ell^C
 W[f_0,g_0](r) + \left(\frac1{a_\ell^C}\right)^{\!2} W[f_2,f_0](r) \\
 - \frac1{a_\ell^C}\left[W[f_2,g_0](r) - W[g_2,f_0](r)\right]
 + W[g_2,g_0](r)\Bigg\} + \OO(p_{A,B}^4) \,.
\label{eq:W_BA-w-exp}
\end{multline}
Note that in the $\mathcal{O}(p_{A,B}^{4})$ we have also included terms of the
form $p_{A}^{2}p_{B}^{2}$.  We set $p_{A}=0$ in Eq.~\eqref{eq:W_BA-w-int} and
furthermore take the limit $p_{B}\rightarrow0$.  Using the
expansion~\eqref{eq:W_BA-w-exp}, we get
\begin{equation}
 {-r_\ell^C}\,W[f_0,g_0](r)
 = b_\ell^C(r) - 2\int_0^r\dd r'\left[w_\ell^{(0)}(r')\right]^2
\label{eq:rC-bC-raw}
\end{equation}
for $r\geq R$, with $w_{\ell}^{(0)}(r)=\lim_{p\rightarrow0}w_{\ell}^{(p)}(r)$
and
\begin{equation}
 b_\ell^C(r) = 2W[g_2,g_0](r) - \frac2{a_\ell^C}\left\{W[f_2,g_0](r)
 + W[g_2,f_0](r)\right\} + \frac2{\left(a_\ell^C\right)^2} W[f_2,f_0](r) \,.
\label{eq:bWronksians}
\end{equation}

\medskip
Written as a function of $\rho=p\cdot r$, the Wronskian of the Coulomb wave
functions is\footnote{See, for example, Eq.~(14.2.4) in
Ref.~\cite{AbramStegPocket}.}
\begin{equation}
 W[F_\ell^{(p)},G_\ell^{(p)}](\rho)
 = -W[G_\ell^{(p)},F_\ell^{(p)}](\rho) = -1 \,.
\end{equation}
Since $\dd/\dd r = p\cdot\dd/\dd\rho$ and $W[f,f]\equiv0$, we also have
\begin{equation}
 W[f,g](r) = W[f,\tilde{g}](r) = -1 \,.
\end{equation}
Plugging in the expansions~\eqref{eq:f-exp} and~\eqref{eq:g-exp}, we see that
$W[f_{0},g_{0}](r)=-1$ for the leading-order functions, and
$W[f_{2},g_{0}](r)=W[g_{2},f_{0}](r)$ for the terms at $\mathcal{O}(p^{2})$.
Inserting these relations into Eq.~\eqref{eq:rC-bC-raw}, we get
\begin{equation}
 r_\ell^C = b_\ell^C(r) - 2\int_0^r\dd r'\left[w_\ell^{(0)}(r')\right]^2 \,,
\label{eq:rC-bC}
\end{equation}
where $b_\ell^C(r)$ has been simplified to
\begin{equation}
 b_\ell^C(r) = 2W[g_2,g_0](r) - \frac4{a_\ell^C}W[f_2,g_0](r)
 + \frac2{\left(a_\ell^C\right)^2}W[f_2,f_0](r) \,.
\label{eq:bC}
\end{equation}
Since the integral in Eq.~\eqref{eq:rC-bC} is positive definite, the resulting
causality bound is
\begin{equation}
 r_\ell^C \leq b_\ell^C(r)\,,\,\forall\ r \geq R \,.
\label{eq:rC-bound}
\end{equation}

\subsection{Calculating the Wronskians}
\label{sec:Wronskians}

We now want to derive the explicit form of the function $b_{\ell}^{C}(r)$.  To
do this, we need expressions for the Wronskians that appear in
Eq.~\eqref{eq:bC}.  We can obtain them by first noting that $f(p,r)$ and
$\tilde{g}(p,r)$, being linear combinations of Coulomb wave functions (with
$p$-dependent coefficients), are solutions of the Coulomb Schr\"{o}dinger
equation,
\begin{equation}
 \left[-\frac{\dd^2}{\dd r^2}+\frac{\ell(\ell+1)}{r^2}
 + \frac{\gamma}r-p^2\right]x(p,r) = 0 \,,
\label{eq:SG-rad-Coulomb}
\end{equation}
which, of course, corresponds to setting $V(r,r^{\prime})=0$ in
Eq.~\eqref{eq:SG-rad-w}.  Here and in the following, $x$ stands for either $f$
or $\tilde{g}$.  Inserting the expansion
\begin{equation}
 x(p,r) = x_0(r) + p^2\,x_2(r) + \OO(p^4)
\label{eq:x-exp}
\end{equation}
into Eq.~\eqref{eq:SG-rad-Coulomb} and comparing orders in $p^{2}$, we find that
\begin{equation}
 \left[-\frac{\dd^2}{\dd r^2}+\frac{\ell(\ell+1)}{r^2}
 + \frac{\gamma}r\right] x_0(r) = 0 \,,
\label{eq:DG-x0}
\end{equation}
\textit{i.e.}, $x_{0}$ is a solution of the zero-energy Coulomb Schr\"{o}dinger
equation, and
\begin{equation}
 \left[-\frac{\dd^2}{\dd r^2}+\frac{\ell(\ell+1)}{r^2}
 + \frac{\gamma}r\right]x_2(r) = x_0(r) \,.
\label{eq:DG-x2}
\end{equation}
From this we readily obtain the differential equations
\begin{subequations}%
\label{eq:DG-Wfg}%
\begin{align}%
\label{eq:DG-Wf2f0}%
 \frac{\dd}{\dd r} W[f_2,f_0](r) &= \left[f_0(r)\right]^2 \,, \\
\label{eq:DG-Wg2g0}%
 \frac{\dd}{\dd r} W[g_2,g_0](r) &= \left[g_0(r)\right]^2 \,, \\
\label{eq:DG-Wf2g0}%
 \frac{\dd}{\dd r} W[f_2,g_0](r) &= f_0(r)g_0(r)
\end{align}%
\end{subequations}%
for the desired Wronskians.  Put together, this yields a simple first-order
differential equation for $b_{\ell}^{C}(r)$,
\begin{equation}
 \frac{\dd}{\dd r}b_{\ell}^{C}(r)
 = 2\left(g_{0}(r)-\frac{1}{a_{\ell}^{C}}f_{0}(r)\right)^{2}.
\label{eq:db}
\end{equation}
From Eqs.~(A.7) and~(A.8) in Ref.~\cite{Bolle:1984ab} we have the explicit
expressions
\begin{subequations}%
\label{eq:f0-g0-rep}%
\begin{align}%
\label{eq:f0-rep}
 f_0(r) &= \frac{(2l+1)!}{\sqrt{\gamma^{2l+1}}}\,\sqrt{r}\,
 I_{2\ell+1}(2\sqrt{\gamma r}) \,, \\
\label{eq:g0-rep}
 g_0(r) &= \frac{2\sqrt{\gamma^{2l+1}}}{(2l+1)!}\,\sqrt{r}\,
 K_{2\ell+1}(2\sqrt{\gamma r})
\end{align}%
\end{subequations}%
for $\gamma>0$, where $I_{\alpha}$ and $K_{\alpha}$ are modified Bessel
functions, and
\begin{subequations}%
\label{eq:f0-g0-attr}%
\begin{align}%
\label{eq:f0-attr}
 f_0(r) &= \frac{(2l+1)!}{\sqrt{(-\gamma)^{2l+1}}}\,\sqrt{r}\,
 J_{2\ell+1}(2\sqrt{-\gamma r}) \,, \\
\label{eq:g0-attr}
 g_0(r) &= \frac{-\pi\sqrt{(-\gamma)^{2l+1}}}{(2l+1)!}\,\sqrt{r}\,
 Y_{2\ell+1}(2\sqrt{-\gamma r})
\end{align}%
\end{subequations}%
for $\gamma<0$, where $J_{\alpha}$ and $Y_{\alpha}$ are the ordinary Bessel
functions.\footnote{From Eq.~(9.1.50) and the remark above Eq.~(9.6.41) in
Ref.~\cite{AbramStegPocket} it is clear that these $f_{0}$ and $g_{0}$ are
indeed solutions of~\eqref{eq:DG-x0}.}\footnote{Boll\'{e} and Gesztesy
actually give an expression for $g_{0}$ in the attractive case ($\gamma<0$)
that involves the Hankel function $H^{(2)}$ times $\mathrm{i}$ instead of $Y$.
With that, however, $g_{0}$ would not be real, which it should be.  Our $g_{0}$
as in Eq.~\eqref{eq:g0-attr} is taken from the results of
Lambert~\cite{Lambert:1968ab}.}

\medskip
Using these expressions for $f_{0}$ and $g_{0}$ and Eq.~\eqref{eq:db} we can
determine $b_{\ell}^{C}(r)$ up to an integration constant.  In order to fix
this constant, we must work directly with the Wronskians in
Eq.~(\ref{eq:bWronksians}).  Before we do that, however, we first discuss the
general form of $b_{\ell}^{C}(r)$.  Let us break apart the function as a sum of
two functions, $X_{\ell}(r)$ and $Y_{\ell}(r)$, and a constant term $Z_{\ell}$,
\begin{equation}
 b_{\ell}^{C}(r) = X_{\ell}(r) + Y_{\ell}(r) + Z_{\ell} \,.
\label{eq:XYZ}
\end{equation}
We take $X_{\ell}(r)$ to be a function consisting entirely of a sum of terms
that have a pole at $r=0$, ranging from order $1$ to $\ell$,
\begin{equation}
 X_{\ell}(r) = \sum\limits_{m=1}^{\ell} X_{\ell,m}r^{-m} \,.
\label{eq:X}
\end{equation}
By furthermore requiring the function $Y_{\ell}(r)$ to vanish at $r=0$, the
decomposition in Eq.~\eqref{eq:XYZ} is unique.

\medskip
Where exactly the contributions to the three terms in the decomposition come
from can be inferred from the behavior of $f(p,r)$ and $g(p,r)$ at the origin.
From Eq.~\eqref{eq:F-rho-zero} combined with Eq.~\eqref{eq:fg-FG} we find
that
\begin{equation}
 f(p,r) \sim r^{\ell+1} \mathtext{as} r\to0 \,.
\end{equation}
This implies that every term in the expansion of $f(p,r)$ is $\OO(r^{\ell+1})$. 
Therefore,
\begin{equation}
 \lim_{r\to0} W[f_2,f_0](r) = 0
\end{equation}
for all $\ell$, which means that this Wronskian only yields contributions to
the $Y_\ell(r)$.

Furthermore, from Eq.~(17) in Ref.~\cite{Yost:1936zz} we know that the
irregular Coulomb wave function has the asymptotic behavior
\begin{equation}
 G_\ell^{(p)} \sim D_{\eta,\ell}\,\rho^{-\ell} \mathtext{as} \rho\to0 \,,
\label{eq:G-rho-to-0} 
\end{equation}
with $D_{\eta,\ell}$ fulfilling $C_{\eta,\ell}D_{\eta,\ell}=2\ell+1$.  Using
Eq.~\eqref{eq:g-G} then yields
\begin{equation}
 g(p,r) \sim \frac{r^{-\ell}}{2\ell+1} \mathtext{as} r\to0 \,.
\end{equation}
We note that $g_{0}(r)$ has exactly the same behavior near $r=0$ and can
thus show that $g_{2}(r)$ is subleading as $r\rightarrow0$, $g_{2}(r)\sim
r^{-\ell+c}$ for $c>0$.  From this we infer that
\begin{equation}
 \lim_{r\to0} W[f_0,g_2](r) = 0
\end{equation}
for all $\ell$, so also from this Wronskian we only get contributions to
$Y_\ell(r)$.  Both the singular $X_\ell(r)$ and the constant $Z_\ell$,
therefore, only come from the Wronskian $W[g_2,g_0](r)$.

For $\ell=0$ the situation is still simple because the above analysis also tells
us that
\begin{equation}
 \lim_{r\to0} W[g_2,g_0](r) = 0 \mathtext{for} \ell=0 \,,
\end{equation}
\ie, $b_{0}^{C}(r)$ is given entirely by $Y_0(r)$.  With the knowledge that it
vanishes at the origin, it is actually straightforward to give an explicit
expression for $b^C_0(r)$ in terms of antiderivatives of the right hand side of
Eq.~\eqref{eq:db}, where one has to insert the $f_0(r)$ and $g_0(r)$ from
Eqs.~\eqref{eq:f0-g0-rep} and~\eqref{eq:f0-g0-attr}.  The result, obtained by
integrating from $0$ to $r$, is
\begin{equation}
\begin{split}
 b_{0}^{C}(r) &=
 \frac{2r^{3}}{3}\left(a_{0}^{C}\right)^{-2}\,
  {_{1}F_{2}}\left(\frac{3}{2};2,4;4\gamma r\right)
 - \frac{4r^{2}}{\sqrt{\pi}}\left(a_{0}^{C}\right)^{-1}
  G_{1,3}^{2,1}\left(4\gamma r\middle|\begin{array}[c]{c}
   {\frac12}\\
    0,1,-2
  \end{array}\right) \\
 &\hspace{2em} + 4\sqrt{\pi}\gamma r^{2}\,
  G_{2,4}^{3,1}\left(4\gamma r\middle|\begin{array}[c]{c}
   -1,{\frac12}\\
   -1,0,1,-2
  \end{array}\right)
\label{eq:b0-rep}
\end{split}
\end{equation}
for the repulsive case, and
\begin{equation}
\begin{split}
 b_{0}^{C}(r) &=
 \frac{2r^{3}}{3}\left(a_{0}^{C}\right)^{-2}\,
  {_{1}F_{2}}\left(\frac{3}{2};2,4;4\gamma r\right)
 + 4\sqrt\pi r^{2}\left(a_{0}^{C}\right)^{-1}
  G^{2,1}_{2,4}\left(-4\gamma r\middle|\begin{array}[c]{c}
   {\frac12},{-\frac12}\\
    0,1,-2,{-\frac12}
  \end{array}\right) \\
 &\hspace{2em} + 2\pi^2\left[\frac{\gamma^2r^3}{3}\,
  {_{1}F_{2}}\left(\frac{3}{2};2,4;4\gamma r\right)
  -\frac{2\gamma r^2}{\sqrt\pi}\,
  G^{3,1}_{3,5}\left(-4\gamma r\middle|\begin{array}[c]{c}
   -1,{\frac12},{-\frac12}\\
    -1,0,1,-2,{-\frac12}
  \end{array}\right)\right]
\label{eq:b0-attr}
\end{split}
\end{equation}
for an attractive Coulomb interaction.  In the above equations,
${_pF_q}$ and $G^{m,n}_{p,q}$ denote the (generalized) hypergeometric and Meijer
$G$-functions, respectively.

\medskip
For general $\ell\geq1$, $W[g_{2},g_{0}](r)$ is singular at $r=0$ and the
analysis becomes more complicated.  For practical purposes, one can simply use
power-series expansions for the Bessel functions that appear in the expressions
for the zero-energy functions and integrate these term by term until a desired
precision is reached.  The only additional ingredients needed are the values for
the constant terms $Z_\ell$ because these are obviously not generated by the
integration.  In Table~\ref{tab:Wg2g0-const} we list these constants for
$\ell=0,1,2$. 

\begin{table}[htbp]
\centering
 \begin{tabular}{c||c|c|c}
  $\;\ell\;$ & $\ 0\ $ & $1$ & $2$ \\
  \hline\hline
  $\,Z_\ell\,$ &
  $\ 0\ $ &
  $\gamma\left(\dfrac{1}{6} - \dfrac{2\gamma_E}{9}\right)$ &
  $\gamma^3\left(\dfrac{79}{21600}-\dfrac{\gamma_E}{360}\right)$
 \end{tabular}
\caption{Constant term $Z_{\ell}$ in Eq.~(\ref{eq:XYZ}) for $\ell=0,1,2$.
$\gamma_{E}=0.577216\ldots$ is the Euler-Mascheroni constant. The values are the
same for repulsive ($\gamma>0$) and attractive ($\gamma<0$) Coulomb potentials.}
\label{tab:Wg2g0-const}
\end{table}

In Appendix~\ref{sec:Wg2g0-const} we describe how the values in
Table~\ref{tab:Wg2g0-const} can be obtained with the help of computer algebra
software.  In that appendix we also give explicit expressions for the causality
bound functions $b_{1}^{C}(r)$ and $b_{2}^{C}(r)$.

\section{Examples and Results}
\label{sec:Results}

\subsection{The causal range}

The causality bound~\eqref{eq:rC-bound} can be rewritten as
\begin{equation}
 b_\ell^C(r) - r_\ell^C \geq 0\ \forall\ r \geq R \,.
\label{eq:rC-bound-2}
\end{equation}
In cases where the details of the interaction (in particular its range,
assuming that a description with finite-range potentials is applicable) is not
known, one can use Eq.~\eqref{eq:rC-bound-2} to define the \emph{causal range}
$R_{c}$ of a scattering system as that value of $r$ for which the bound is just
satisfied, \ie,
\begin{equation}
 b_\ell^C(R_c) - r_\ell^C = 0 \,.
\label{eq:rC-bound-CR}
\end{equation}
We note from Eq.~\eqref{eq:db} that the derivative of $b_{\ell}^{C}(r)$ is
non-negative,
\begin{equation}
 \frac{\dd}{\dd r} b_\ell^C(r)
 = 2\left(g_0(r)-\frac1{a_\ell^C}f_0(r)\right)^2 \geq 0 \,.
\label{eq:db-positive}
\end{equation}
Hence, $b_{\ell}^{C}$ is an increasing function of $r$ and the causal range is
defined uniquely.  For the case that Eq.~\eqref{eq:rC-bound-CR} does not have a
solution (\ie, if $b_\ell^C(r)$ is positive already for $r=0$), we define the
causal range to be zero.  Note that the causal range is a function only of the
scattering length and effective range.  It can thus be calculated from
observables in a well-defined way.

The importance of the causal range is given by the
fact that it can be interpreted as the minimum range a potential is allowed to
have to be consistent with causality.  If for a given system the values in
individual partial waves differ significantly, the maximum value should be taken
as the causal range of the underlying potential.  Alternatively, one can model
the interaction with an $\ell$-dependent potential.  For effective field
theories with short-range interactions such as halo EFT, the causal range
constrains the allowed values of the momentum-space cutoff or the lattice
spacing used in numerical calculations. 

\medskip
At this point we recall that our derivation of the causality bounds was based on
the assumption that the concrete system under consideration is described by a
finite-range (though possibly non-local) two-body interaction which is energy
independent.  In EFT calculations one frequently obtains effective interactions
that explicitly depend on the energy.  We note that this energy dependence can
be traded for momentum dependence at any given order in the power counting (EFT
expansion) by using the equations of motion obtained from the effective
Lagrangian.  However, the energy dependence introduces another length scale into
the system, and so the conversion to momentum-dependent interactions could
produce an interaction range so large that the causality bounds may not be
useful in practice.

There are also other theoretical frameworks, e.g., Feshbach reaction theory,
that explicitly use energy-dependent interactions.  Here again the energy
dependence introduces a length scale which acts as an interaction range.  This
can be seen from the time delay of the scattered wavepacket, which is
proportional to the derivative of the phase shift with respect to energy.  By
setting a very strong energy dependence for the interactions, it is possible to
produce a time delay which is arbitrarily large and negative.  This has the same
effect as interactions at arbitrarily large separation distances.

Furthermore, the assumption of a strict finite range certainly is an
idealization that is only applicable to a varying degree of validity to concrete
physical systems.  For example, there can be exchange forces arising from the
Pauli principle.  Consider for example nuclear halo systems with a tightly bound
core and a halo nucleon which is only weakly bound to the core.  The exchange of
a nucleon from the core and the halo nucleon that is necessary to
anti-symmetrize the system can only give a sizeable contribution if there is
spatial overlap between the wavefunction of the core and the wavefunction of the
halo nucleon.  This yields a short-range exponential tail that, within the
domain of validity of the effective theory, can be subsumed in the effective
range parameters of the halo-core interaction.  The same analysis would apply to
low-energy nucleon-nucleus scattering upon the core nucleus.  Another more
prominent effect is given by exponential tails generated by simple pion-exchange
contributions; see Ref.~\cite{Elhatisari:2012ym} and the discussion below.

\begin{figure}[ptbh]
\centering
\includegraphics[width=0.6\textwidth,clip]{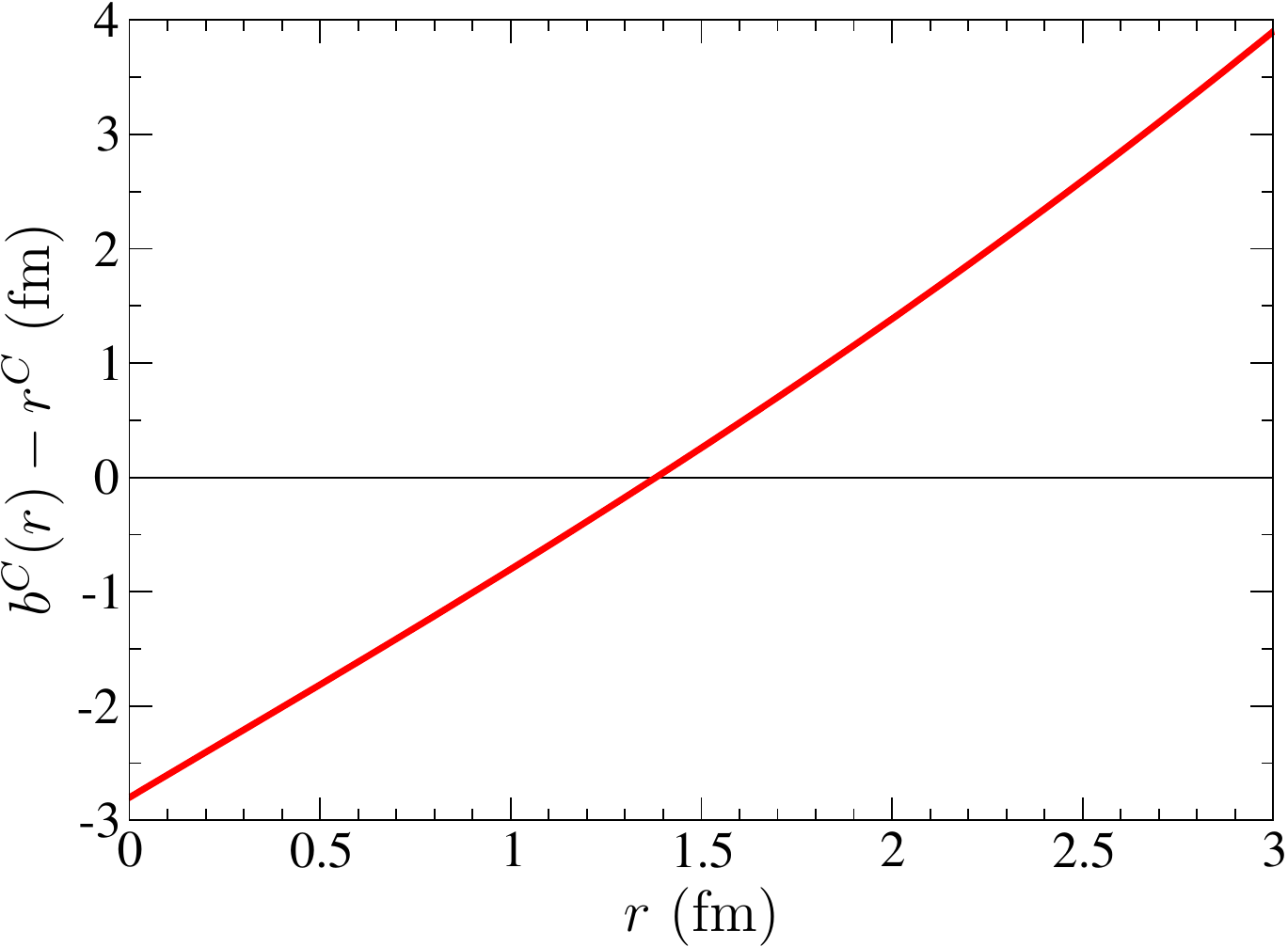}
\caption{Causal range plot for S-wave proton--proton scattering.}%
\label{fig:Range-pp-0}%
\end{figure}

\begin{figure}[ptbh]
\centering
\includegraphics[width=0.65\textwidth,clip]{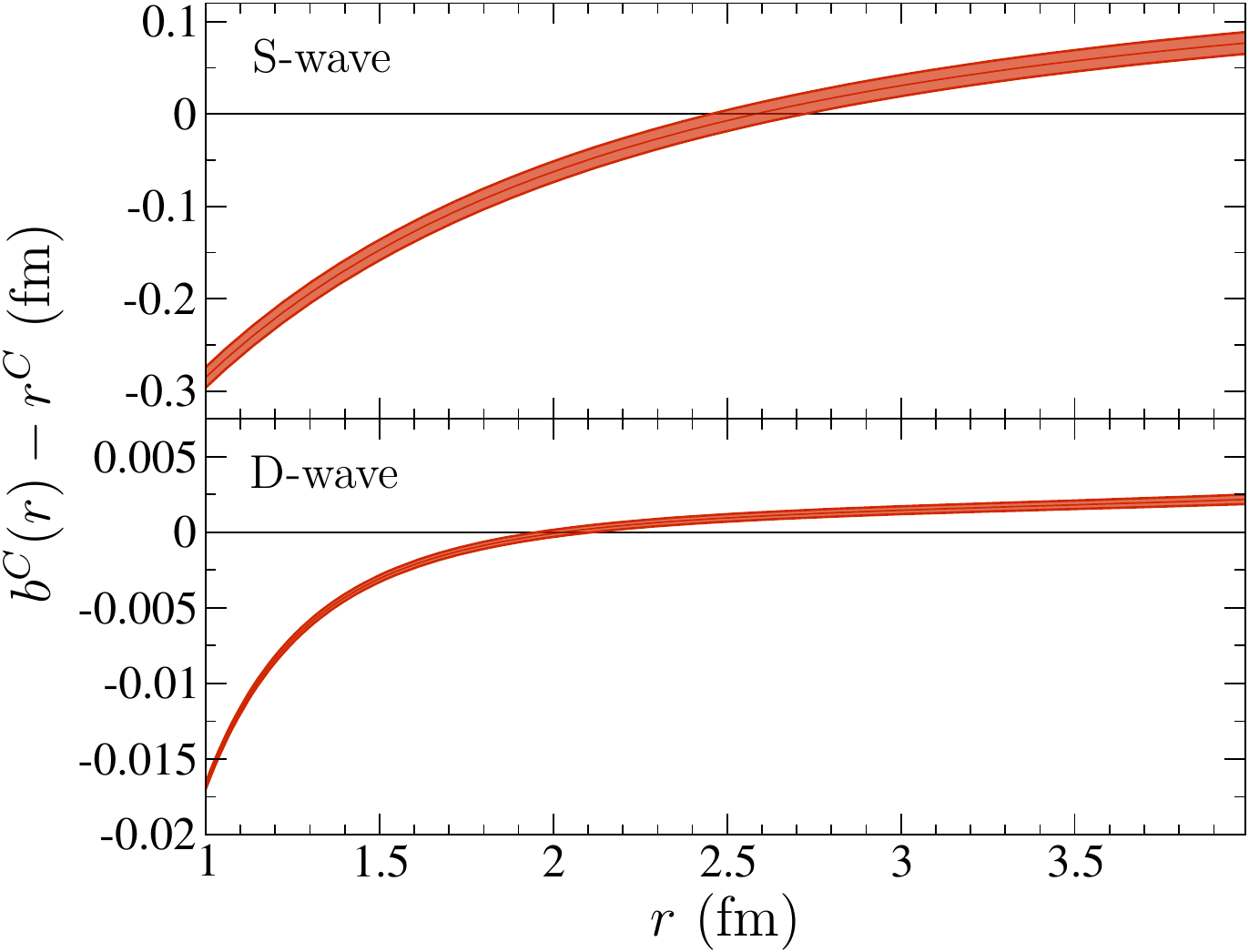}
\caption{Causal range plot for S-wave $\alpha$--$\alpha$ scattering.}%
\label{fig:Range-aa}%
\end{figure}

\medskip
We now calculate explicit values for causal ranges in few-nucleon systems.  In
Fig.~\ref{fig:Range-pp-0} we plot the left-hand side of
Eq.~\eqref{eq:rC-bound-2} as a function of $r$ for the case of proton--proton
S-wave scattering.  The causal range can then be read off as the point where
the function becomes zero.  Fig.~\ref{fig:Range-aa} shows analogous plots for
$\alpha$--$\alpha$ S- and D-wave scattering.  In this system, there are visible
error bands due to the larger uncertainties in the effective range parameters.

In Table~\ref{tab:Results} we give a summary of the causal ranges that one
finds for various two-body systems of light nuclei where low-energy scattering
parameters and/or phase shifts are available from experiments.  The results are
briefly discussed in the following subsections.

\begin{table}[htbp]
\centering
 \begin{tabular}{c|c|c|c|c|c}
  System & Reference & Channel & $a^C /\ \fm^{2\ell+1}$
  & $r^C /\ \fm^{-2\ell+1}$ & Causal range / $\fm$ \\
  \hline\hline
  $p$--$p$ & \cite{Naisse:1977ab} & $^1S_0$
  & $-7.828\pm0.008$ & $2.80\pm0.02$ & $1.38\pm0.01$ \\
  $p$--$p$ & \cite{Bergervoet:1988zz} & $^3P_0$
  & $-3.03\pm0.11$ & $4.22\pm0.11$ & $2.33\pm0.05$ \\
  $p$--$p$ & \cite{Bergervoet:1988zz} & $^3P_1$
  & $2.013\pm0.053$ & $-7.92\pm0.17$ & $\approx0.03$ \\
  \hline
  $p$--$d$ & \cite{Arvieux:1973ab} & $^2S_{1/2}$
  & $2.73\pm0.10$ & $2.27\pm0.12$ & $3.90\pm0.15$ \\
  $p$--$d$ & \cite{Huttel:1983ab} & $^2S_{1/2}$
  & $4$ & $-2.8$ & $0$ \\ 
  $p$--$d$ & \cite{Arvieux:1973ab} & $^4S_{3/2}$
  & $11.88\pm0.40$ & $2.63\pm0.02$ & $2.20^{+0.07}_{-0.06}$ \\
  $p$--$d$ & \cite{Huttel:1983ab} & $^4S_{3/2}$
  & $11.11$ & $2.64$ & $2.29$ \\
  \hline
  $p$--$^3\mathrm{He}$ & \cite{Daniels:2010af} & $^1S_{0}$
  & $11.1\pm0.4$ & $1.58\pm0.12$ & $1.32^{+0.21}_{-0.17}$ \\
  $p$--$^3\mathrm{He}$ & \cite{Daniels:2010af} & $^3S_{1}$
  & $9.04\pm0.14$ & $1.50\pm0.06$ & $1.27^{+0.10}_{-0.09}$ \\  
  \hline
  $p$--$\alpha$ & \cite{Arndt:1973ab} & $S_{1/2}$
  & $4.97\pm0.12$ & $1.295\pm0.082$ & $1.32^{+0.40}_{-0.21}$ \\
  $p$--$\alpha$ & \cite{Arndt:1973ab} & $P_{1/2}$
  & $-19.36\pm0.50$ & $0.349\pm0.021$ & $2.65\pm0.07$ \\
  $p$--$\alpha$ & \cite{Arndt:1973ab} & $P_{3/2}$
  & $-44.83\pm0.51$ & $-0.365\pm0.113$ & $0.49^{+0.17}_{-0.10}$ \\
  \hline
  $\alpha$--$\alpha$ & \cite{Higa:2008dn} & $S$
  & $(-1.65\pm0.17)\cdot10^3$ & $1.084\pm0.011$ & $2.58^{+0.14}_{-0.13}$ \\
  $\alpha$--$\alpha$ & \cite{Afzal:1969zz} & $D$
  & $(-7.23\pm0.61)\cdot10^3$ & $(-1.31\pm0.22)\cdot10^{-3}$
  & $2.03^{+0.09}_{-0.07}$
 \end{tabular}
\caption{Summary of causal-range results obtained from experimental input for
various few-nucleon systems.}
\label{tab:Results}
\end{table}

\subsubsection{Proton--proton scattering}
\label{sec:Results-pp}

For $p$--$p$ S-wave scattering one finds a causal range of about
$1.38~\mathrm{fm}$.  This value is very close to the range estimate
obtained by assuming that the typical length scale of the $N$--$N$
interaction is set by the inverse pion mass, $\hbar c/M_{\pi}\approx1.4~\fm$. 
The value one finds in the $^{3}P_{0}$-channel is somewhat larger
($R_{c}\approx2.3~\fm$), whereas the $^{3}P_{1}$ effective range parameters
impose almost no constraint on the range of the nuclear potential in this
channel.  As we will discuss in more detail below, this suggests some
significant differences in the radial dependence of the interactions for the
$^{3}P_{1}$ channel.

For effective field theory calculations with purely local interactions (\ie,
pionless effective field theory), our results suggest to keep the cutoff
momentum smaller than $M_\pi$ for the $^{1}S_{0}$ and  $^{3}P_{0}$ channel. 
However, there is more freedom to take a higher cutoff for the $^{3}P_{1}$
channel.

In Ref.~\cite{Elhatisari:2012ym}, causality bounds were investigated for
neutron--proton scattering.  The results ($R_{c}=1.27~\fm$ for
$^{1}S_{0}$, $R_{c}=3.07~\fm$ for $^{3}P_{0}$, and $R_{c}=0.23~\fm$ for
$^{3}P_{1}$) are qualitatively very similar to what we find for the $p$--$p$
system.  This would indicate only a moderate amount of isospin breaking.

In the same publication~\cite{Elhatisari:2012ym}, the influence of the shape of
the potential upon the neutron--proton causal range was also studied
numerically. When the potential is repulsive at shorter distances (less
than $\sim1~\fm$) and attractive at larger distances (greater than $\sim1~\fm$),
the causal range comes out on the larger side, about $2~\fm$ or more.  When
the potential is attractive at intermediate distances and repulsive at larger
distances, then the causal range is smaller, about $1~\fm$ or less.  The pion
tail determines the sign of the potential at larger distances.  For both the
$n$--$p$ and the $p$--$p$ interaction, the one-pion exchange tail is repulsive
in the $^{3}P_{1}$-channel while it is attractive in the $^{3}P_{0}$-channel.

Note that causality bounds in the presence of pion-exchange contributions were
also discussed by Phillips and Cohen in Ref.~\cite{Phillips:1996ae}.  Ideally,
one would account for the one-pion exchange tail explicitly in the calculation
of the causal range, as it was done in this work for the long-range Coulomb
potential.  Without knowing analytical solutions of the Schrödinger equation
involving a Yukawa-like potential (plus a Coulomb part, in the $p$--$p$ case),
however, such a procedure can at best be implemented numerically.  For an
example, see Ref.~\cite{Scaldeferri:1996nx}.

\subsubsection{Proton--deuteron scattering}
\label{sec:Results-pd}

There are several experimental determinations of $p$--$d$ effective range
parameters. In Table~\ref{tab:Results}, we have included results from
Arvieux~\cite{Arvieux:1973ab} and Huttel~\textit{et al.}~\cite{Huttel:1983ab}.
While for the quartet-channel there is a good agreement between the scattering
lengths and effective ranges (and, of course, of the resulting causal ranges,
which come out as $2.2-2.3~\fm$), there is a large discrepancy for
the doublet-channel results.

The difficulty of determining the proton--deuteron doublet-channel scattering
length has previously been discussed by Orlov and Orevkov~\cite{Orlov:2006ab}. 
Comparing different models, the authors conclude that $a^{C}=0.024~\fm$ is
currently the best theoretical estimate for the doublet-channel $p$--$d$
scattering length.  From Table 3 in Ref.~\cite{Orlov:2006ab} one reads off that
the corresponding value for the Coulomb-modified effective range is as huge as
$r^{C}=8.23\cdot10^{5}~\fm$.  Inserting these numbers into the causal-range
calculation one gets a very large value of $R_{c}\approx8.15~\fm$.  As a
consequence, no definite conclusion can be reached with the currently available
data.

Note that three-body forces have been found to be very important for theoretical
calculations of the $p$--$d$ (and $n$--$d$) threshold scattering parameters. 
Our analysis here, however, is independent of the microscopic origin of the
effective interaction between proton and deuteron.  In a detailed picture, the
force might arise from two-nucleon forces or three-nucleon forces, but the
result is always some effective two-body interaction between the proton and the
deuteron. The causal range we calculate is the minimum range that this effective
interaction has to have in order to be able to reproduce the experimentally
determined scattering parameters.

\subsubsection{Proton--helion scattering}
\label{sec:Results-p3He-pa}

For the scattering of protons off a helium nucleus we were able to find data
for both $p$--$^3\mathrm{He}$ and $p$--$\alpha$ scattering.  In the first case,
there was only enough data available to calculate the causal range for the
S-wave channels.  Since both the scattering lengths and effective ranges are
very similar for the singlet and the triplet channel, so are the resulting
causal ranges, which come out as approximately $1.3~\fm$.

Incidentally, one finds almost the same value for the S-wave in $p$--$\alpha$
scattering.  For this system, it is interesting to compare to the neutron--alpha
system, where there is no Coulomb repulsion in the scattering process.  Results
for the $n$--$\alpha$ causal ranges can be read off from Fig.~5 in
Ref.~\cite{Hammer:2010fw} (obtained using effective range parameters from 
Ref.~\cite{Arndt:1973ab}).  Even though from the plot one only gets quite rough
estimates, one clearly sees that the results for the $S_{1/2}$ and $P_{1/2}$
channels agree very well between $p$--$\alpha$ and $n$--$\alpha$ scattering,
which as in the nucleon--nucleon case discussed above could be interpreted as
only a moderate amount of isospin breaking.  However, the causal ranges for the
$P_{3/2}$ channels are very different ($\sim0.5~\fm$ for $p$--$\alpha$,
$\sim2~\fm$ for $n$--$\alpha$).  It is an interesting question if this
discrepancy hints at an error in the extraction of the effective range
parameters (either for one of the systems or possibly both), or if there
actually is a physical reason behind the difference in the causal ranges.

\subsubsection{Alpha--alpha scattering}
\label{sec:Results-aa}

For $\alpha$--$\alpha$ S-wave scattering we use the values given by
Higa~\textit{et al.} (see Ref.~\cite{Higa:2008dn} and experimental references
therein), $a_{0}^{C}=(-1.65\pm0.17)\cdot10^{3}~\fm$ and
$r_{0}^{C}=(-1.084\pm0.011)~\fm$ to find a causal range of about $2.58~\fm$.

For the $\ell=2$ channel, no effective range parameters could be found in the
literature.  We have thus used the phase shift data collected in the review
article by Afzal~\textit{et al.}~\cite{Afzal:1969zz} to perform the fit to the
effective range expansion (Eq.~\eqref{eq:ERE-CbMod} with $\ell=2$) ourselves. 
By including the phase-shift data up to $E_{\text{lab}}\approx6.5~\MeV$ we find
$a_{2}^{C}=(-7.23\pm0.61)\cdot10^{3}~\fm^{5}$ and
$r_{2}^{C}=(-1.31\pm0.22)\cdot10^{-3}~\fm^{-3}$.  However, the fit is strongly
dominated by the $\mathcal{O}(p^{4})$ shape parameter, so the actual
uncertainties of $a_{2}^{C} $ and $r_{2}^{C}$ should probably be somewhat
larger.  For the causal range in this channel we find a value of about $2~\fm$,
which is just slightly smaller than the S-wave result.

\section{Relation for asymptotic normalization constants}
\label{sec:ANC-relations}

We now discuss the connection between the integral relations we have derived
above and asymptotic normalization constants (ANC), denoted in the following
as $A$.  They are defined by writing the bound-state wave function in the
asymptotic region as
\begin{equation}
 w_\ell(r) = A\cdot W_{-\ii\eta,\ell+\frac12}(2\kappa r) \mathtext{for} r>R \,,
\label{eq:ANC-def}
\end{equation}
where $w_{\ell}(r)$ is a bound-state solution (with momentum $p=\ii\kappa$,
$\kappa>0$) of the Schr\"{o}dinger equation~\eqref{eq:SG-rad-w}, normalized
according to
\begin{equation}
 \int_0^\infty \dd r\,|w_\ell(r)|^2 = 1 \,.
\label{eq:NC-rad}
\end{equation}
$W_{-\ii\eta,\ell+\frac12}$ is the Whittaker function defined in
Eq.~\eqref{eq:WhittakerW}.  At very low-energies, the radiative capture
cross section to a two-body halo nucleus is proportional to $|A|^2$.

\subsection{Derivation of the relation}

Our starting point is Eq.~\eqref{eq:rC-bC} derived in
Sec.~\ref{sec:CausalityBounds}, from which we already obtained the causality
bounds for the effective range parameter.  For bound-state momenta, we write it
as
\begin{equation}
 r_\ell^C = b_\ell^C(r) - 2 \lim_{\kappa\to0}
 \int_0^r\dd r'\left[w_\ell^{(\ii\kappa)}(r')\right]^2
 \mathtext{for all} r>R \,,
\label{eq:rC-bC-BS}
\end{equation}
with $b_{\ell}^{C}(r)$ as defined in Eq.~\eqref{eq:bC}.  In order to derive
from this a relation for the ANC, we need the precise connection between the
radial wave functions $w_{\ell}^{(\ii\kappa)}$ and the solution
$w_{\ell}(r)$ used in Eq.~\eqref{eq:ANC-def}.  Formally, we can write
\begin{equation}
 w_\ell^{(\ii\kappa)}(r) = (\ii\kappa)^\ell C_{\eta,\ell}
 \left[\cot\tilde{\delta}_\ell(\ii\kappa)\,F_\ell^{(\ii\kappa)}(r)
 + G_\ell^{(\ii\kappa)}(r)\right] \mathtext{for} r>R \,,
\label{eq:w-asympt-BS}
\end{equation}
but we need to be careful with the way the analytic continuation to the
bound-state regime is done.  We start with the assumption that $\cot
\tilde{\delta}_{\ell}(\ii\kappa) = \ii$ for a bound state and thus consider the
linear combination $G_{\ell}^{(p)}(r)+\ii F_{\ell}^{(p)}(r)$, not explicitly
setting $p=\ii\kappa$ for the moment.  A short calculation shows that
\begin{equation}
 G_\ell^{(p)}(r)+\ii F_\ell^{(p)}(r)
 = \ee^{\ii\sigma_\ell}\ee^{-\ii\frac{\pi}2(\ell+\ii\eta)}\,
 W_{-\ii\eta,\ell+\frac12}(-2\ii pr)
\end{equation}
with the Coulomb phase shift $\sigma_{\ell}$ defined
via~\cite{Dzieciol:1999ab,Lambert:1968ab}
\begin{equation}
 \ee^{\ii\sigma_\ell} = \left(\frac{\Gamma(\ell+1+\ii\eta)}
 {\Gamma(\ell+1-\ii\eta)}\right)^{\!\frac12} \,.
\label{eq:sigma-ell}
\end{equation}
Note that $\ii\eta$ will be real for bound-state momenta.  In the following, we
only consider the case of a repulsive Coulomb potential (\ie, $\gamma>0$), so
$\ii\eta$ is in fact a positive number.  For the $C_{\eta,\ell}$, we use the
expression~\cite{Humblet:1984ab}
\begin{equation}
 C_{\eta,\ell} = \frac{2^\ell\,\ee^{-\frac{\pi\eta}2}
 \left[\Gamma(\ell+1+\ii\eta)\Gamma(\ell+1-\ii\eta)\right]^{\frac12}}
 {\Gamma(2\ell+2)} \,,
\label{eq:C_eta-complex-gen}
\end{equation}
\cf~also Eq.~\eqref{eq:C_eta-complex-BG} in Appendix~\ref{sec:CoulombWF-BG}. 
Combining Eqs.~\eqref{eq:sigma-ell} and~\eqref{eq:C_eta-complex-gen} we find
that the problematic terms $\Gamma(\ell+1-\ii\eta)$ and $\exp(\pi\eta/2)$ drop
out, and we can finally write
\begin{equation}
 w_\ell^{(\ii\kappa)}(r) = \kappa^\ell\tilde{C}_{\eta,\ell}\,
 W_{-\ii\eta,\ell+\frac12}(2\kappa r) \mathtext{for} r>R
\end{equation}
where we have defined
\begin{equation}
 \tilde{C}_{\eta,\ell}
 = \frac{2^{\ell}\,\Gamma(\ell+1+\ii\eta)}{\Gamma(2\ell+2)} \,.
\label{eq:C-tilde}
\end{equation}
Comparing with Eq.~\eqref{eq:ANC-def} we readily infer that
\begin{equation}
 w_\ell^{(\ii\kappa)}(r)
 = \frac{\kappa^\ell}{A}\tilde{C}_{\eta,\ell}\cdot w_\ell(r) \,.
\end{equation}
Using the overall normalization of the wave function, we can rewrite
Eq.~\eqref{eq:rC-bC-BS} as
\begin{equation}
 r_\ell^C = b_\ell^C(r) - 2 \lim_{\kappa\to0} \left\{
 \frac{\kappa^{2\ell}}{A^2}\tilde{C}_{\eta,\ell}^2
 - \int_r^\infty\dd r'\left[w_\ell^{(\ii\kappa)}(r')\right]^2 \right\}
 \mathtext{,} r>R \,.
\label{eq:rC-bC-BS-2}
\end{equation}

\subsubsection*{Canceling the $r$-dependence}
At this point it is important to note that Eq.~\eqref{eq:rC-bC-BS-2} can be
rigorously valid only for zero-energy bound states, \ie, if one strictly
considers the limit $\kappa\to0$.  Since the left hand side is a constant, the
$r$-dependence of $b_\ell^C(r)$ has to cancel that of the integral for any
$r>R$.  In the scattering regime this is ensured by taking the zero-energy
limit, \ie, considering scattering directly at threshold.  In the bound-state
regime, on the other hand, one is of course interested in finding a relation for
the case where there is a bound state close to threshold, \ie, where $\kappa$ is
small but finite.  It is obvious that in principle this poses a problem
because the $b_\ell^C(r)$ in Eq.~\eqref{eq:rC-bC-BS-2} is a strictly increasing
function of $r$, whereas the integral is always bounded and in fact becomes
smaller when $r$ is made larger.

\medskip
However, up to corrections of higher order in $\kappa$, the cancellation still
works, \ie, the \emph{leading} $r$-dependence drops out.  In
Ref.~\cite{Koenig:2011ti}, where we derived an analogous ANC relation for the
case without any long-range force, we established the cancellation by carrying
out the integral over the asymptotic wave function (a Riccati-Hankel function,
in that case) analytically and then expanding the result in powers of
$\kappa$.  To show it here in a more general way, we recall from
Eq.~\eqref{eq:w-asympt-ERE} that
\begin{equation}
 w_\ell^{(p)}(r) = -\frac1{a_\ell^C}f_0(r) + g_0(r) + \OO(p^2) \,.
\label{eq:w-asympt-ERE-short} 
\end{equation}
It is straightforward to assume that this is valid also in the bound-state
regime since we are working with a wave function explicitly analytic in $p^2$. 
Hence,
\begin{equation}
 \left[w_\ell^{(\kappa)}(r)\right]^2
 = \left(-\frac1{a_\ell^C}f_0(r) + g_0(r)\right)^2 + \OO(\kappa^2)
 = \frac12\frac{\dd}{\dd r}b_{\ell}^{C}(r) + \OO(\kappa^2) \,, 
\end{equation}
where in the last step we have inserted Eq.~\eqref{eq:db}.  This directly tells
us that $b_{\ell}^{C}(r)$ in Eq.~\eqref{eq:rC-bC-BS-2} cancels with the
integral up to higher-order terms and, possibly, an integration constant.  This
situation is already familiar from the calculation of the Wronskian
$W[g_2,g_0](r)$.  Again the constant term can only come from the integral over
$g_0(r)^2$.  To determine it, we note that
\begin{equation}
 \int_r^\infty\dd r'\left[g_0(r')\right]^2 < \infty \mathtext{if} r>0 \,,
\end{equation}
\ie, the integral is actually convergent and gives a Meijer $G$-function, as
already encountered in Sec.~\ref{sec:Wronskians}.  The constant term coming from
this function has to be combined with the already known $Z_\ell$ in
$b_{\ell}^{C}(r)$.  We define
\begin{equation}
 \Delta Z_{\ell} = W[g_2,g_0](r) + \int_r^\infty\dd r'\left[g_0(r')\right]^2\,,
\end{equation}
which is explicitly $r$-independent, to finally arrive at
\begin{equation}
 r_\ell^C + \frac{2\kappa^{2\ell}}{A^2}\tilde{C}_{\eta,\ell}^2
 + 2\,\Delta Z_{\ell} = \OO(\kappa^2)
\label{eq:ANC-rel}
\end{equation}
or, equivalently,
\begin{equation}
 |A| = \frac{\kappa^\ell}{\tilde{C}_{\eta,\ell}}
 \left(-\frac{r_\ell^C}{2}-\Delta Z_{\ell} + \OO(\kappa^2)\right)^{-1/2} \,,
\label{eq:ANC-rel-solved}
\end{equation}
both valid in the limit where $\kappa\to0$.  In Table~\ref{tab:cg2g0} we give
the resulting values for the $\Delta Z_{\ell}$ for $\ell=0,1,2$.

\begin{table}[htbp]
\centering
 \begin{tabular}{c||c|c|c}
  $\;\ell\;$ & $\ 0\ $ & $1$ & $2$ \\
  \hline\hline
  $\,\Delta Z_\ell,$ & $-1/{(3\gamma)}$ & $-\gamma/108$ & $-17\gamma^3/10800$
\end{tabular}
\caption{Integration constant $\Delta Z_{\ell}$ 
in Eqs.~(\ref{eq:ANC-rel}, \ref{eq:ANC-rel-solved}) for $\ell=0,1,2$.}
\label{tab:cg2g0}
\end{table}

\medskip
Finally, we note that Sparenberg \etal~\cite{Sparenberg:2009rv} have previously
derived an ANC relation equivalent to ours, but written in terms of the
scattering length instead of the effective range parameter.  The equivalence of
the two relations up to the contributions of higher order shape parameters is
shown in Appendix \ref{sec:Sparenberg-equiv}. 

\subsection{Application to the oxygen-16 system}

The $^{16}\mathrm{O}$ nucleus has two excited states lying just below the
$\alpha$--$^{12}\mathrm{C}$ threshold, a $2^+$ at about $-245~\keV$ and an even
more shallow $1^-$ at only $-45~\keV$.  The properties of these states play an
important role in astrophysical helium burning
processes~\cite{Brune:1999zz,Wallerstein:1997ab,Buchmann:1996zz}.
In the following, we calculate asymptotic normalization constants for both
states under the assumption that they can be described in a
$\alpha+{^{12}\mathrm{C}}$ halo picture.

We use the recent data obtained by Tischhauser~\etal~\cite{Tischhauser:2009zz}
(for the actual phase shifts see their Ref.~[32]) in order to extract the
Coulomb-modified effective range for the $\alpha$--$^{12}\mathrm{C}$ P- and
D-wave channels.  Focusing first on the D-wave, we note that the combination of
the strong Coulomb repulsion and the $\ell=2$ centrifugal barrier makes the
low-energy phase shifts very small over a wide energy range.  Moreover, there is
a narrow resonance at a center-of-mass energy of about $2.7~\MeV$, which
strongly constrains the energy region for a straightforward fit to the effective
range function.  From a simple fit to the data up to the narrow resonance one
only obtains an effective range parameter with an uncertainty too large (50\%)
for an extraction of the ANC since the latter depends on $r^C$ very sensitively.

To mitigate this problem, we use the position of the $2^+$ oxygen state as an
additional input parameter for a self-consistent extraction of the effective
range.  In the following, we describe this procedure in more detail.

\medskip
At the position of the bound state, where $\cot\tilde{\delta}_{\ell}(\ii\kappa)
= \ii$, the Coulomb-modified effective range
expansion~\eqref{eq:ERE-CbMod-complex} reduces to
\begin{equation}
 \gamma\,\tilde{h}_\ell(\ii\kappa)
 = -\frac1{a^C_\ell} - \frac12r^C_\ell\,\kappa^2 + \cdots \,,
\label{eq:ERE-CbMod-complex-BS}
\end{equation}
where now
\begin{equation}
 \tilde{h}_\ell(\ii\kappa) = \frac{(2\ii\kappa)^{2\ell}}{(2\ell+1)!^2}
 \prod_{s=1}^\ell(s^2+\eta^2)
 \times\left(\psi(\ii\eta)+\frac1{2\ii\eta}-\log(\ii\eta)\right)
 \mathtext{,} \eta=\frac{\gamma}{2\ii\kappa} \,.
\label{eq:h-tilde-BS}
\end{equation}
A straightforward calculation shows that for the prefactor we have
\begin{equation}
 \frac{(2\ii\kappa)^{2\ell}}{(2\ell+1)!^2}\prod_{s=1}^\ell(s^2+\eta^2)
 = \frac{\gamma^{2\ell}}{(2\ell+1)!^2} + \OO(\kappa^2) \,.
\end{equation}
Furthermore, for the digamma function we have the asymptotic expansion
(Eq.~(6.3.18) in Ref.~\cite{AbramStegPocket})
\begin{equation}
 \psi(z) = \log z - \frac1{2z}
 - \sum\limits_{n=1}^\infty\frac{B_{2n}}{2n\,z^{2n}}
\label{eq:psi-exp}
\end{equation}
for $|z|\to\infty$, where the $B_{2n}$ are the Bernoulli numbers,
\begin{equation}
 B_0 = 1,\ B_1=-\frac12,\ B_2=\frac16,\ \ldots\ \,.
\end{equation}
Noting that the sum in Eq.~\eqref{eq:psi-exp} only starts at $n=1$, we find that
$\gamma\,\tilde{h}_\ell(\ii\kappa)$ only starts at order $\kappa^2$, and
inserting the precise relation into Eq.~\eqref{eq:ERE-CbMod-complex-BS}, we
arrive at
\begin{equation}
 \frac1{a^C_\ell} =\left(-\frac12r^C_\ell
 + \frac{\gamma^{2\ell-1}}{3(2\ell+1)^2}\right)\kappa^2
 + \OO(\kappa^4) \,.
\label{eq:aC-kappa}
\end{equation}

Neglecting the effective range contribution at leading order, we insert
the binding momentum $\kappa\approx0.187~\fm^{-1}$ of the $2^+$ state to get a
first approximation for the scattering length parameter.  This is then used to
constrain a subsequent fit to the phase shift data up to about $2.6~\MeV$,
\ie, just below the narrow resonance.  We include a single $\OO(p^4)$ shape
parameter in the fit since some curvature is clearly necessary.  The $r^C_2$
obtained from this is then used in Eq.~\eqref{eq:aC-kappa} to get a better
result for $1/a^C_2$, which, in turn, is fed back into the fit.  Iterating this
procedure a couple of times yields a well-converged self-consistent result for
$r^C_2$.  After eight iterations, we find
\begin{equation}
 r^C_2 = (5.94\pm0.35)\cdot10^{-4}~\fm^{-3}
\end{equation}
for the $\alpha$--$^{12}\mathrm{C}$ D-wave.  Including a second shape parameter
in the fitting procedure only changes this result within the given uncertainty,
so we conclude that for the energy range we have been fitting, a single shape
parameter really is sufficient to account for the curvature.  Inserting the
fit result into Eq.~\eqref{eq:ANC-rel-solved} yields
\begin{equation}
 |A(2^+)| = (2.41\pm 0.38)\cdot 10^{4}~\fm^{-1/2}
\end{equation}
for the $2^+$ state in $^{16}\mathrm{O}$.  Including in Eq.~\eqref{eq:ANC-rel}
an $\OO(\kappa^4)$ term of the order of the shape parameter gives a consistent
result for $|A(2^+)|$ within the error given above.

\medskip
When we apply the same same procedure to the $1^-$ state just $45~\keV$ below
the $\alpha$--$^{12}\mathrm{C}$ threshold, we also see a nice convergence and
obtain the results
\begin{equation}
 r^C_1 = (4.546\pm0.002)\cdot10^{-2}~\fm^{-1}
\end{equation}
and
\begin{equation}
 |A(1^-)| = (1.188\pm0.024)\cdot10^{14}~\fm^{-1/2} \,.
\end{equation}
The overall picture for the $1^-$ state is somewhat more complicated, however. 
On the one hand, the extraction of the effective range parameters is easier in
this case because there is no narrow resonance limiting the fit range (we have
used the data up to $E_{\mathrm{cm}}=3.75~\MeV$).  By doing a simple fit without
the self-consistent iteration we get values for the effective range and the ANC
that are slightly smaller, but overlap with the results given above when the
respective uncertainties are taken into account.  On the other hand, allowing
for a second shape parameter changes the value of the effective range quite
dramatically to $0.046~\fm^{-1}$, which leads to an imaginary ANC.  This could
indicate that the effective range parameters from the simple fit violate the
causality bound.  The fact that $r^C$ has to be such that Eq.~\eqref{eq:ANC-rel}
yields a real value for the ANC can be interpreted as a weaker remnant of the
original causality bound relation.  Alternatively, the cluster picture might not
be applicable for the shallow $1^-$ state in $^{16}\mathrm{O}$.

\medskip
Our result for the ANC of the $2^+$ state is about a factor five smaller
than the value $|A(2^+)| = (1.11\pm 0.11)\cdot 10^{5}~\fm^{-1/2}$ obtained by
Brune \etal~\cite{Brune:1999zz}, while the value for $|A(1^-)|$ is about a
factor two smaller.  Other, more recent
determinations~\cite{Adhikari:2011zz,Belhout:2007ab} have found even larger
values.  For the interpretation of this discrepancy, note that our calculations
are predictions of the ANCs based only on alpha--carbon elastic scattering data
and the assumption that the system can be approximately described in an
effective two-body picture with a finite-range interaction.  In the references
mentioned above, the ANCs are extracted from alpha--carbon transfer
measurements.  Note furthermore that a comparison of the experimental
extractions in Ref.~\cite{Adhikari:2011zz} exhibits quite some discrepancy
(factors of two up to roughly an order of magnitude) also between the cited
individual experimental determinations of the ANCs.

Sparenberg~\etal~\cite{Sparenberg:2011zz} have carried out a similar
analysis for the $2^+$ state based on their ANC relation
(\cf~Ref.~\cite{Sparenberg:2009rv}).  They subtracted the narrow resonance from
the $\alpha$--$^{12}\mathrm{C}$ D-wave phase shift data in order to extract a
set of higher-order shape parameters and concluded that present day data are not
sufficient to constrain the ANC strongly.  As discussed above, our approach of
performing a self-consistent fit to the data below the resonance constrained by
the separation energy and using the effective range instead of the scattering
length as input in the ANC relation improves the stability of the extraction.
However, compared to other determinations our ANC values are generally smaller
by a factor two to five.  We thus conclude that this issue requires further
study.

\section{Conclusion}

In this paper, we have investigated the constraints imposed by causality on the
low-energy scattering parameters of charged particles interacting via a
short-range interaction and a long-range Coulomb potential.  Similar to the
case of neutral particles without Coulomb
interaction~\cite{Phillips:1996ae,PavonValderrama:2005wv,Cordon:2009pj,
Cordon:2009wh,Hammer:2009zh,Hammer:2010fw,Elhatisari:2012ym}, our
considerations lead to a constraint on the maximum value of the Coulomb-modified
effective range. 

While conceptionally straightforward, the calculation of the Wronskians of the
wave functions required for the derivation of the bound is intricate.  We have
calculated them through term-by-term integration of the power series expansion
of the zero-energy wave functions and additionally determining the integration
constants that are not generated by this process.

We define the causal range as the minimum value of the interaction range 
consistent with the causality bound.  In effective field theories with contact
interactions such as the halo EFT, the natural momentum cutoff is of the order
of the inverse of the causal range.  If the natural cutoff is not known from
other considerations, its size  can be estimated from the causal range.  If the
momentum cutoff used in a calculation is too high, then problems with
convergence of higher order corrections can appear.  For example, the
convergence pattern might be such that an improvement in higher orders of
the EFT can only be sustained through large cancellations between individual
terms.  Such an unnatural pattern would be especially undesirable for the
stability of numerical (lattice) calculations.  Our results can thus be viewed
as a guide for improving the convergence pattern of EFT calculations with
contact interactions. In lattice simulations of halo EFT, the lattice spacing
should not be taken smaller than the causal range.

We have analyzed the causal ranges for a variety of systems ranging from 
proton--proton scattering to alpha--alpha scattering.  Our results for causal 
ranges in different partial waves in these systems typically vary by factors of
2-3.  The precise values are quite sensitive to small uncertainties in the 
effective range parameters.  In channels with a large negative effective range
the causal range is very close to zero, which implies that causality provides
almost no constraints on the range of the interaction in this case.  Thus the
causal range provides a good order of magnitude estimate of the range of
interaction, but drawing more quantitative conclusions about the structure of
the underlying potentials is difficult.

After an analytic continuation to the bound state regime, the integral relations
for the causality bound can also be used to derive a model-independent
expression for the ANC for shallow bound states.  If the state is a two-body
halo state, the relation can be used to extract the ANC from low-energy
scattering parameters.  Up to higher order shape parameters, our relation is
equivalent to the one previously derived by Sparenberg \etal~in
Ref.~\cite{Sparenberg:2009rv} (see Appendix \ref{sec:Sparenberg-equiv} for
details).  One difference is that we express the ANC in terms of the binding
momentum and the effective range rather than the binding momentum and the
scattering length.  We find this form more suitable for the extraction of ANCs
from scattering data since the effective range is typically more precisely
determined than the scattering length for shallow states.  Moreover, extracting
the effective range in a self-consistent fit from the scattering data that
reproduces the correct separation energy improves the stability of the
extraction.

Finally, we illustrate our relation by extracting the ANCs of the excited $2^+$
and $1^-$ states in $^{16}$O from $\alpha-^{12}$C scattering data.  Compared to
previous extractions~\cite{Brune:1999zz,Adhikari:2011zz,Belhout:2007ab}, our
values are generally smaller.  Whether this difference is physically significant
requires further study.  The application of our relation to other shallow
cluster states and a benchmark against model calculations would also be very
interesting.

\begin{acknowledgments}
We thank Renato Higa and Gautam Rupak for useful discussions.  We also thank the
Institute for Nuclear Theory at the University of Washington for its hospitality
and the Department of Energy for partial support during the completion of this
work.  S.K. furthermore thanks the North Carolina State University, where a part
of this research was worked out, for its hospitality.  This research was
supported in part by the DFG through SFB/TR 16 ``Subnuclear structure of
matter,'' the BMBF under contracts No. 06BN9006, 06BN7008, and by the US
Department of Energy under contract No. DE-FG02-03ER41260.  S.K. was supported
by the ``Studien\-stiftung des deutschen Volkes'' and by the Bonn-Cologne
Graduate School of Physics and Astronomy.
\end{acknowledgments}

\appendix

\section{Calculating the constant terms in the causality bound function}
\label{sec:Wg2g0-const}

To determine the constants $Z_{\ell}$ introduced in Sec.~\ref{sec:Wronskians},
we consider the explicit form of $\tilde{g}(p,r)$.  From the results of
Boll\'{e} and Gesztesy~\cite{Bolle:1984ab}, we have\footnote{One gets this form
from Eq.~\eqref{eq:G-tilde-raw} in Appendix~\ref{sec:CoulombWF-BG} by inserting
$n=2\ell+3$, $\eta=\gamma/(2p)$ and assuming that the momentum $p$ is real and
positive.}
\begin{multline}
 \tilde{g}(p,r) = \mathcal{N}_\ell(p)\cdot\gamma\log\left(|\gamma|r\right)
 \cdot f(p,r) \\
 + \gamma\,\Rp\left\{\mathcal{N}_\ell(p)\cdot r^{\ell+1}\,\ee^{-\ii pr}
 \cdot\sum\limits_{n=0}^\infty\left[a_{\ell,n}(p)+b_{\ell,n}(p)\right]
 r^n\right\}  \\
 + \Rp\left\{\frac{r^{-\ell}}{2\ell+1}\,\ee^{-\ii pr}
 \cdot\sum\limits_{n=0}^{2\ell} d_{\ell,n}(p)\,r^n\right\} \,,
\label{eq:g-tilde-series}
\end{multline}
where
\begin{equation}
 \mathcal{N}_\ell(p) = \frac{(2p)^{2\ell}}{\Gamma(2\ell+2)^2}
 \prod\limits_{s=1}^\ell(s^2+\eta^2) \,,
\end{equation}
\begin{subequations}
\begin{equation}
 a_{\ell,n}(p) = \frac{-\Gamma(2l+2)}{\Gamma(n+1)\Gamma(n+2\ell+2)}(2\ii p)^n
 \prod_{s=1}^n(s+\ell-\ii\eta)\cdot\big[\psi(n+1)+\psi(n+2\ell+2)\big] \,,
\end{equation}
\begin{equation}
 b_{\ell,n}(p) = \frac{\Gamma(2l+2)}{\Gamma(n+1)\Gamma(n+2\ell+2)}(2\ii p)^n
 \prod_{s=1}^n(s+\ell-\ii\eta)
 \cdot\sum\limits_{j=1}^{n+\ell}\frac1{j-\ii\eta} \,,
\end{equation}
\end{subequations}
and
\begin{equation}
 d_{\ell,n}(p) = \frac1{\Gamma(n+1)}(2\ii p)^n
 \prod\limits_{s=1}^n\left(\frac{s-\ell-1-\ii\eta}{s-2\ell-1}\right) \,.
\end{equation}

With this result and the appropriate expression for $g_0(r)$ from
Eq.~\eqref{eq:f0-g0-rep} or ~\eqref{eq:f0-g0-attr}, one can use the following
procedure to calculate the $Z_{\ell}$, \ie, the terms of order $r^0$ in the
Wronskian $W[g_2,g_0](r)$.
\begin{enumerate}
 \item Note that
 \begin{equation}
  W[\tilde{g},g_{0}](r) = p^{2}\,W[g_{2},g_{0}](r) + \OO(p^{4})
 \end{equation}
 and calculate this Wronskian using a truncated version (including terms up to
 the order $2\ell+1$ in $r$ is sufficient) of $\tilde{g}(p,r)$ as given in
 Eq.~\eqref{eq:g-tilde-series}.
 \item From the result, extract the terms that are of the order $r^0$.
 \item From that expression then extract the terms that are of the order
 $p^2$.  They constitute the $\OO(r^0)$-contributions in a series expansion of
 $W[g_2,g_0](r)$ which cannot be obtained from a term-by-term integration of
 $g_0(r)^2$.
\end{enumerate}
With a computer algebra software, this prescription is straightforward to
implement.  The results for $\ell=0,\ldots,2$ are shown in
Table~\ref{tab:Wg2g0-const} in Sec.~\ref{sec:Wronskians}.

\medskip
At this point we remark that there is a recent publication by
Seaton~\cite{Seaton:2002ab} discussing Coulomb wave functions that are
explicitly analytic in the energy (or the momentum squared).  In principle, 
it is possible to use these results to get explicit expressions for $f_2(r)$ and
$g_2(r)$ (in addition to the already known zero-energy functions), and then
simply calculate all the Wronskians directly.  However, the analytic irregular
Coulomb function defined by Seaton is slightly different from our
$\tilde{g}(p,r)$. More importantly, not all coefficients needed for the 
expansions in energy are given explicitly.  Finally, 
writing everything in terms of
the wave functions of Boll\'{e} and Gesztesy paves the way for a generalization
of the results presented in the current paper to an arbitrary number of spatial
dimensions.

\medskip
Knowing now what the constant terms $Z_\ell$ that come from $W[g_2,g_0](r)$
should be, it is possible to write down explicit expressions for the causality
bound functions $b^C_\ell(r)$.  To do that we use that the antiderivative of the
right hand side of Eq.~\eqref{eq:db} can be expressed in terms of (generalized)
hypergeometric functions ${_pF_q}$ and Meijer $G$-functions $G^{m,n}_{p,q}$. 
The only additional point to be taken into account is that the antiderivative of
$g_0(r)^2$ in general includes a constant term that is different from the
desired $Z_\ell$.  Hence, one has to determine this term and add another
constant such that the sum is exactly equal to the $Z_\ell$ given in
Table~\ref{tab:Wg2g0-const}.  For the antiderivative of $g_0(r)^2$ that we are
using in the following, up to a minus sign the correction term is just the
$\Delta Z_\ell$ introduced in Sec.~\ref{sec:ANC-relations}.\footnote{In other
words, our choice for the antiderivative corresponds to using $\int_r^\infty\dd
r'g_0(r')^2$, at least for the case of a repulsive Coulomb interaction, where
this integral is convergent.}

\subsubsection{Repulsive case, $\gamma>0$}

The $\ell=0$ result for a repulsive Coulomb potential is given by
Eq.~\eqref{eq:b0-rep}.  For $\ell=1$, we get
\begin{equation}
\begin{split}
 b_{1}^{C}(r) &= \frac{6r^2}{5\gamma^3}\left(a_{1}^{C}\right)^{-2}\left[
  3\gamma^{2}r^{2}\,{_1F_2}\left(\frac52;3,6;4\gamma r\right)
  - 20\gamma r\,{_1F_2}\left(\frac32;2,5;4\gamma r\right)\right. \\
  &\hspace{8em}\left. + 120\,{_1F_2}\left(\frac12;1,4;4\gamma r\right)
  - 120\,{_2F_3}\left(\frac12,2;1,3,4;4\gamma r\right)\right] \\
 &\hspace{2em} - \frac{4r^2}{\sqrt\pi}\left(a_{1}^{C}\right)^{-1}
  G^{2,2}_{2,4}\left(4\gamma r\middle|\begin{array}[c]{c}
   -1,\frac{1}{2}\\
   0,3,-3,-2
  \end{array}\right) \\
 &\hspace{2em} - \frac{\sqrt{\pi}\gamma^{3}r^{2}}{9}\,G^{4,0}_{2,4}\left(
  4\gamma r\middle|\begin{array}[c]{c}
   -1,\frac{1}{2}\\
   -3,-2,0,3
  \end{array}\right) + \frac{\gamma}{54} \,,
\end{split}
\label{eq:b1-rep}
\end{equation}
and for $\ell=2$ the result is
\begin{equation}
\begin{split}
 b_{2}^{C}(r) &= \frac{50r^2}{63\gamma^5}\left(a_{2}^{C}\right)^{-2}\left[
  7\gamma^{4}r^{4}\,{_1F_2}\left(\frac92;5,10;4\gamma r\right)
  - 108\gamma^3r^3\,{_1F_2}\left(\frac72;4,9;4\gamma r\right)\right. \\
  &\hspace{8em}\left.+ 1296\gamma^2r^2\,{_1F_2}
  \left(\frac52;3,8;4\gamma r\right)
  - 12096\gamma r\,{_1F_2}\left(\frac32;2,7;4\gamma r\right) \right. \\
  &\hspace{8em}\left. + 108864\,{_1F_2}\left(\frac12;1,6;4\gamma r\right)
  - 108864\,{_2F_3}\left(\frac12,2;1,3,6;4\gamma r\right)\right] \\
 &\hspace{2em}- \frac{4r^2}{\sqrt\pi}\left(a_{2}^{C}\right)^{-1}
  G^{2,2}_{2,4}\left(4\gamma r\middle|\begin{array}[c]{c}
   -1,\frac{1}{2}\\
   0,5,-5,-2
  \end{array}\right) \\
 &\hspace{2em}- \frac{\sqrt{\pi}\gamma^{5}r^{2}}{3600}\,G^{4,0}_{2,4}\left(
  4\gamma r\middle|\begin{array}[c]{c}
   -1,\frac{1}{2}\\
   -5,-2,0,5
  \end{array}\right) + \frac{\gamma^3}{21600} \,.
\end{split}
\label{eq:b2-rep}
\end{equation}

\subsubsection{Attractive case, $\gamma<0$}

For an attractive Coulomb potential, the $\ell=0$ result is given by
Eq.~\eqref{eq:b0-attr}.  The $\ell=1$ result reads
\begin{equation}
\begin{split}
 b_{1}^{C}(r) &= \frac{6r^2}{5\gamma^3}\left(a_{1}^{C}\right)^{-2}\left[
  3\gamma^{2}r^{2}\,{_1F_2}\left(\frac52;3,6;4\gamma r\right)
  - 20\gamma r\,{_1F_2}\left(\frac32;2,5;4\gamma r\right)\right. \\
  &\hspace{8em}\left. + 120\,{_1F_2}\left(\frac12;1,4;4\gamma r\right)
  - 120\,{_2F_3}\left(\frac12,2;1,3,4;4\gamma r\right)\right] \\
 &\hspace{2em}+ 4\sqrt\pi r^2\left(a_{1}^{C}\right)^{-1} G^{2,2}_{3,5}\left(
  -4\gamma r\middle|\begin{array}[c]{c}
   -1,\frac12,{-\frac12}\\
   0,3,-3,-2,{-\frac12}
  \end{array}\right) \\
 &\hspace{2em}+ \frac{\pi^{3/2}\gamma^3r^2}{18}\left[2\,G^{4,0}_{3,5}\left(
  -4\gamma r\middle|\begin{array}[c]{c}
   {-\frac52},-1,{\frac12}\\
   -3,-2,0,3,{-\frac52}
  \end{array}\right)\right. \\
 &\hspace{9em}\left.- G^{1,2}_{2,4}\left(
  -4\gamma r\middle|\begin{array}[c]{c}
   -1,{\frac12}\\
   3,-3,-2,0
  \end{array}\right)\right] + \frac{\gamma}{54} \,,
\end{split}
\label{eq:b1-attr}
\end{equation}
and for $\ell=2$ one finds
\begin{equation}
\begin{split}
 b_{2}^{C}(r) &= \frac{50r^2}{63\gamma^5}\left(a_{2}^{C}\right)^{-2}\left[
  7\gamma^{4}r^{4}\,{_1F_2}\left(\frac92;5,10;4\gamma r\right)
  - 108\gamma^3r^3\,{_1F_2}\left(\frac72;4,9;4\gamma r\right)\right. \\
  &\hspace{8em}\left. + 1296\gamma^2r^2\,{_1F_2}
  \left(\frac52;3,8;4\gamma r\right)
  - 12096\gamma r\,{_1F_2}\left(\frac32;2,7;4\gamma r\right) \right. \\
  &\hspace{8em}\left. + 108864\,{_1F_2}\left(\frac12;1,6;4\gamma r\right)
  - 108864\,{_2F_3}\left(\frac12,2;1,3,6;4\gamma r\right)\right] \\
 &\hspace{2em}+ 4\sqrt\pi r^2\left(a_{2}^{C}\right)^{-1}G^{2,2}_{3,5}\left(
  -4\gamma r\middle|\begin{array}[c]{c}
   -1,{\frac12},{-\frac12}\\
   0,5,-5,-2,{-\frac12}
  \end{array}\right) \\
 &\hspace{2em} + \frac{\pi^{3/2}\gamma^{5}r^{2}}{7200}\left[2\,
 G^{4,0}_{3,5}\left(-4\gamma r\middle|\begin{array}[c]{c}
   {-\frac92},-1,{\frac12}\\
   -5,-2,0,5,{-\frac92}
  \end{array}\right)\right. \\
  &\hspace{9em}\left.- G^{1,2}_{2,4}\left(
  -4\gamma r\middle|\begin{array}[c]{c}
   -1,{\frac12}\\
   5,-5,-2,0
  \end{array}\right)\right] + \frac{\gamma^3}{21600} \,.
\end{split}
\label{eq:b2-attr}
\end{equation}

\section{The Coulomb wave functions of Boll\'{e} and Gesztesy}
\label{sec:CoulombWF-BG}

Boll\'{e} and Gesztesy define the Coulomb wave functions
\begin{subequations}%
\begin{equation}
 F_n^{(0)}(p,r) = r^{\frac12+m} \ee^{-\ii pr}
 {_1F_1}\!\left(\tfrac12+m-\kappa,1+2m;z\right)
\label{eq:F-n}
\end{equation}
and
\begin{equation}
 G_n^{(0)}(p,r) = \frac{\Gamma\left(\frac12+m-\kappa\right)}{\Gamma(2m+1)}
 (2\ii p)^{2m}\,r^{\frac12+m}\ee^{-\ii pr}
 U\!\left(\tfrac12+m-\kappa,1+2m;z\right) \,,
\label{eq:G-n}
\end{equation}
\label{eq:FG-n}%
\end{subequations}%
where, in our case, $n=2\ell+3$ and $m$, $\kappa$, $z$ are as defined in
Eq.~\eqref{eq:rho-z-kappa-m}. Using the formula\footnote{From Eq.~(3.1) in
Ref.~\cite{Boersma:1969ab} one directly sees that this is consistent with
Eq.~\eqref{eq:C_eta-complex} in Sec.~\ref{sec:ERE}.}
\begin{equation}
 C_{\eta,\ell} = \frac{2^\ell\ee^{-\pi\eta/2}\,|\Gamma(\ell+1+\ii\eta)|}
 {(2\ell+1)!} \,,
\label{eq:C_eta-complex-BG}
\end{equation}
Eq.~(5.1) in Ref.~\cite{Boersma:1969ab}, one finds that
\begin{subequations}%
\begin{equation}
 F_n^{(0)}(p,r) = \frac{1}{p^{\ell+1}C_{\eta,\ell}} F_\ell^{(p)}(r)
\end{equation}
and
\begin{equation}
 G_n^{(0)}(p,r) = p^{\ell}C_{\eta,\ell}
 \left[G_\ell^{(p)}(r)-\ii F_\ell^{(p)}(r)\right] \,.
\end{equation}
\label{eq:FG-n-FG}%
\end{subequations}%
It is shown in Ref.~\cite{Bolle:1984ab} that $F_{n}^{(0)}(p,r)$ is analytic in
$p^{2}$. Furthermore, from Eqs.~(3.16), (3.17) and~(4.1) in that paper it
follows that
\begin{equation}
 G_n^{(0)}(p,r) = \tilde{G}_n^{(0)}(p,r)
 + \left(\gamma\,\tilde{h}_\ell(p) - \ii p^{2\ell+1} C_{\eta,\ell}^2\right)
 \cdot F_n^{(0)}(p,r) \,,
\end{equation}
where $\tilde{h}_{\ell}(p)$ is the function defined in Eq.~\eqref{eq:h-tilde},
and where $\tilde{G}_{n}^{(0)}(p,r)$ is analytic in $p^{2}$.  The functions
$f(p,r)$ and $\tilde{g}(p,r)$ that we introduce in Sec.~\ref{sec:RewriteWF}
are exactly the analytic wave functions defined above, \ie,
\begin{equation}
 f(p,r) = F_n^{(0)}(p,r) \mathtext{and}
 \tilde{g}(p,r) = \tilde{G}_n^{(0)}(p,r) \,.
\end{equation}
It should be noted that Lambert~\cite{Lambert:1968ab} already defines Coulomb
wave functions that are analytic in $p^{2}$, but only for the most common case
of three spatial dimensions. Boll\'{e} and Gesztesy extend Lambert's results
to an arbitrary number of dimensions $\geq2$.

\medskip
More importantly, at least for our application, in Eq.~(4.3) of
Ref.~\cite{Bolle:1984ab}, Boll\'{e} and Gesztesy give an explicit expression for
$\tilde{G}_{n}^{(0)}(p,r)$.  Since there are two typos in their
original equation,\footnote{The $\mathrm{i}\gamma/k$ in the first line should be
$\mathrm{i}\gamma/(2k) $ and the $(q+1)$ in the last line should be
$\Gamma(q+1)$.} we quote the whole expression for completeness.  
Slightly altering the
notation to match our conventions, we have
\begin{multline}
 \tilde{G}_n^{(0)}(p,r) = \left[\Gamma(n-1)^{-2}\right](2p)^{n-3}
 \left|\Gamma\left(\frac{n-1}2+\frac{\ii\gamma}{2p}\right)\right|^2
 \left|\Gamma\left(1+\frac{\ii\gamma}{2p}\right)\right|^{-2}
 \gamma\log\left(|\gamma|r\right)\cdot F_n^{(0)}(p,r) \\
 - \gamma\,\mathrm{Re}\Bigg\{\left[\Gamma(n-1)^{-2}\right](2p)^{n-3}
 \left|\Gamma\left(\frac{n-1}2+\frac{\ii\gamma}{2p}\right)\right|^2
 \left|\Gamma\left(1+\frac{\ii\gamma}{2p}\right)\right|^{-2}r^{(n-1)/2}
 \,\ee^{-\ii pr} \\
 \times\sum\limits_{k=0}^\infty\frac{\Gamma\left(\frac{n-1}2+k
 -\frac{\ii\gamma}{2p}\right)\Gamma(n-1)(2\ii pr)^k}
 {\Gamma\left(\frac{n-1}2-\frac{\ii\gamma}{2p}\right)
 \Gamma(n-1+k)\Gamma(k+1)}\left[\psi(k+1)+\psi(k+n+1)\right]\Bigg\} \\
 + \gamma\,\mathrm{Re}\Bigg\{\left[\Gamma(n-1)^{-2}\right](2p)^{n-3}
 \left|\Gamma\left(\frac{n-1}2+\frac{\ii\gamma}{2p}\right)\right|^2
 \left|\Gamma\left(1+\frac{\ii\gamma}{2p}\right)\right|^{-2}r^{(n-1)/2}
 \,\ee^{-\ii pr} \\
 \times\sum\limits_{k=0}^\infty\frac{\Gamma\left(\frac{n-1}2+k
 -\frac{\ii\gamma}{2p}\right)\Gamma(n-1)(2\ii pr)^k}
 {\Gamma\left(\frac{n-1}2-\frac{\ii\gamma}{2p}\right)\Gamma(n-1+k)\Gamma(k+1)}
 \sum\limits_{s=1}^{\frac{n-1}2+k-1}\left(s-\frac{\ii\gamma}{2p}\right)^{-1}
 \Bigg\} \\
 + \mathrm{Re}\Bigg\{(n-2)^{-1}(2\ii p)^{n-2}r^{(n-1)/2}\,\ee^{-\ii pr}
 \sum\limits_{q=0}^{n-3}\frac{\Gamma\left(\frac{3-n}2+q-\frac{\ii\gamma}{
 2p}\right)\Gamma(3-n)(2\ii pr)^{q+2-n}}
 {\Gamma\left(\frac{3-n}2-\frac{\ii\gamma}{2p}\right)\Gamma(q+3-n)\Gamma(q+1)}
 \Bigg\} \,,
\label{eq:G-tilde-raw}
\end{multline}
valid for any odd $n\geq3$ and where the sum over $s$ in the fourth line is
defined to give zero for $n=3$ and $k=0$.

\section{Numerical calculations}
\label{sec:Numerics}

In order to check our relations and to get a better understanding of the values
for the causal range, we consider some explicit examples numerically.

By cutting off the singular parts of the potential (\ie, the Coulomb potential
and the angular momentum term for $\ell\geq0$) at very small distances, it is a
simple task to numerically solve the radial Schr{\"o}dinger
equation~\eqref{eq:SG-rad-w} in configuration space.  From the radial
wave functions one can extract the Coulomb-modified phase shifts by looking for
a zero at some large (\ie, much larger than the range $R$ of the short-range
potential) distance,
\begin{equation}
 w^{(p)}_\ell(r_0) = 0 \mathtext{,} r_0 \gg R \,,
\end{equation}
and then calculating
\begin{equation}
 \cot\tilde{\delta}_\ell(p) = -\frac{G_\ell{(p)}(r_0)}{F_\ell{(p)}(r_0)} \,.
\end{equation}
For the simplest case of a local step potential,
\begin{equation}
 V(r,r') = V_\text{step}(r)\cdot\delta(r-r')
 \equiv V_0\,\theta(R-r)\cdot\delta(r-r') \,,
\end{equation}
one can of course also obtain the phase shift directly by matching the wave
functions at $r=R$.  The effective range parameters are then obtained by
repeating the calculation for several (small) momenta and fitting
Eq.~\eqref{eq:ERE-CbMod} to the results.

In order to test Eq.~\eqref{eq:rC-bC} directly one needs the wave function to
calculate the integral
\begin{equation}
 \int_0^r\dd r'\left[w_\ell^{(0)}(r')\right]^2 
 = \lim_{p\to0}\int_0^r\dd r'\left[w_\ell^{(p)}(r')\right]^2 \,.
\end{equation}
Even if we do not actually take the limit $p\to0$ but rather just insert some
small $p_{0}=0.1$ (in units of an arbitrary inverse length scale), we find
that the relation
\begin{equation}
 r_\ell^C
 \approx b_\ell^C(R) - 2\int_0^R\dd r'\left[w_\ell^{(p_0)}(r')\right]^2 \,,
 \label{eq:rC-bC-approx}
\end{equation}
is typically fulfilled to better than one percent accuracy for the simple step
potential defined above.

\medskip
For illustration, in the following we choose units where the radial distance is
measured in $\fm$.  The potential range is set to $1~\fm$ and its strength is
measured in $\mathrm{MeV}$.\footnote{Note that with these conventions, the
quantity that is used in the numerical calculation is $v_{0} \equiv 2\mu
V_{0}/(\hbar c)^{2}$, where $\hbar c \approx197.33~\MeV\cdot\fm$ is used for the
unit conversion.}  Furthermore, the reduced mass and Coulomb parameter are set
to the values for the proton--proton system, \ie, $2\mu = m_{N} \approx
940~\MeV$ and $\gamma = \gamma_{\text{p--p}} \approx0.035~\fm^{-1}$.

\begin{figure}[htbp]
\centering
\includegraphics[width=0.62\textwidth,clip]{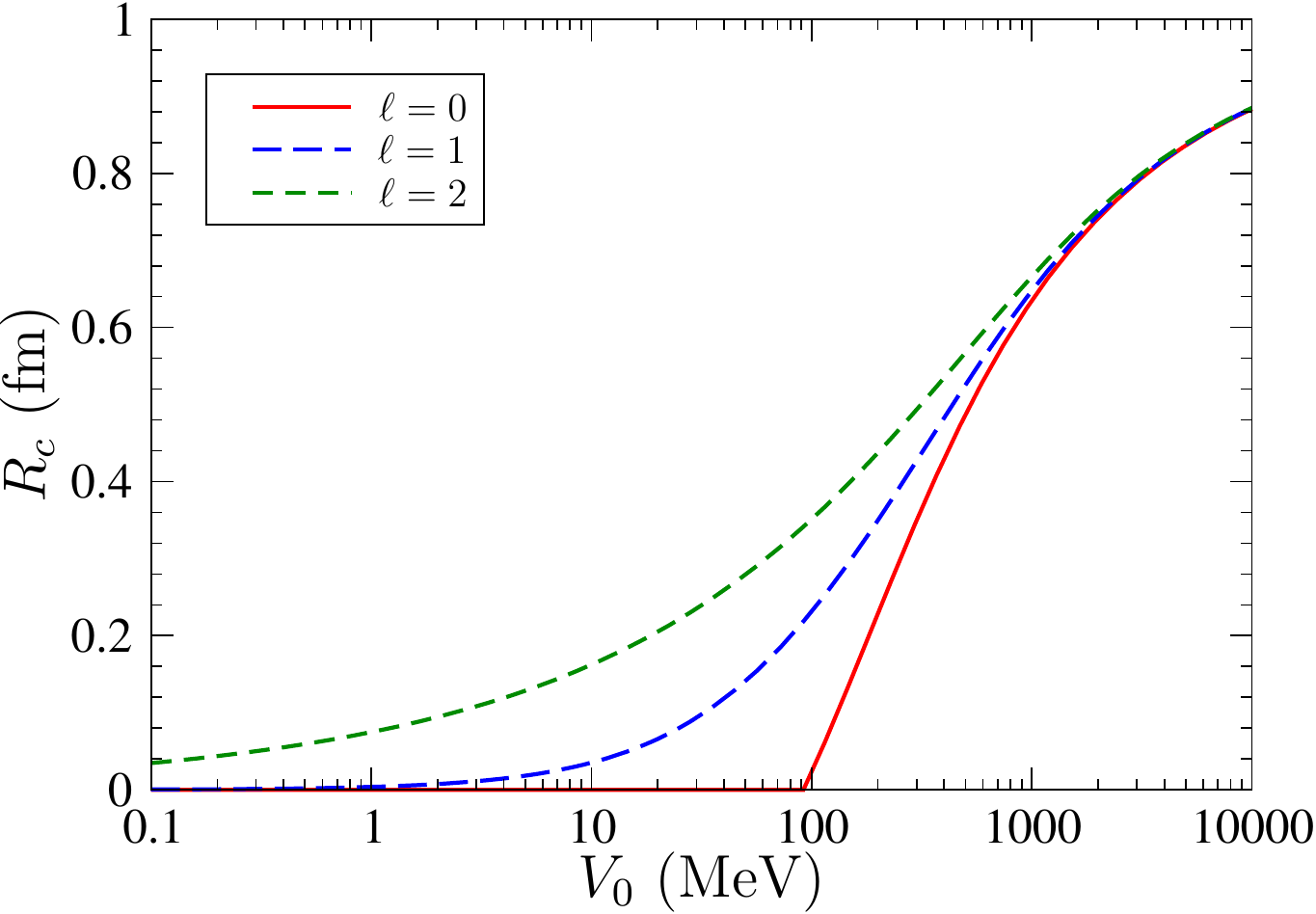}
\caption{Causal range for a repulsive step potential and $\gamma =
\gamma_{\text{p--p}}$.}
\label{fig:Rc-step-rep-gpp}
\end{figure}
\begin{figure}[htbp]
\centering
\includegraphics[width=0.62\textwidth,clip]{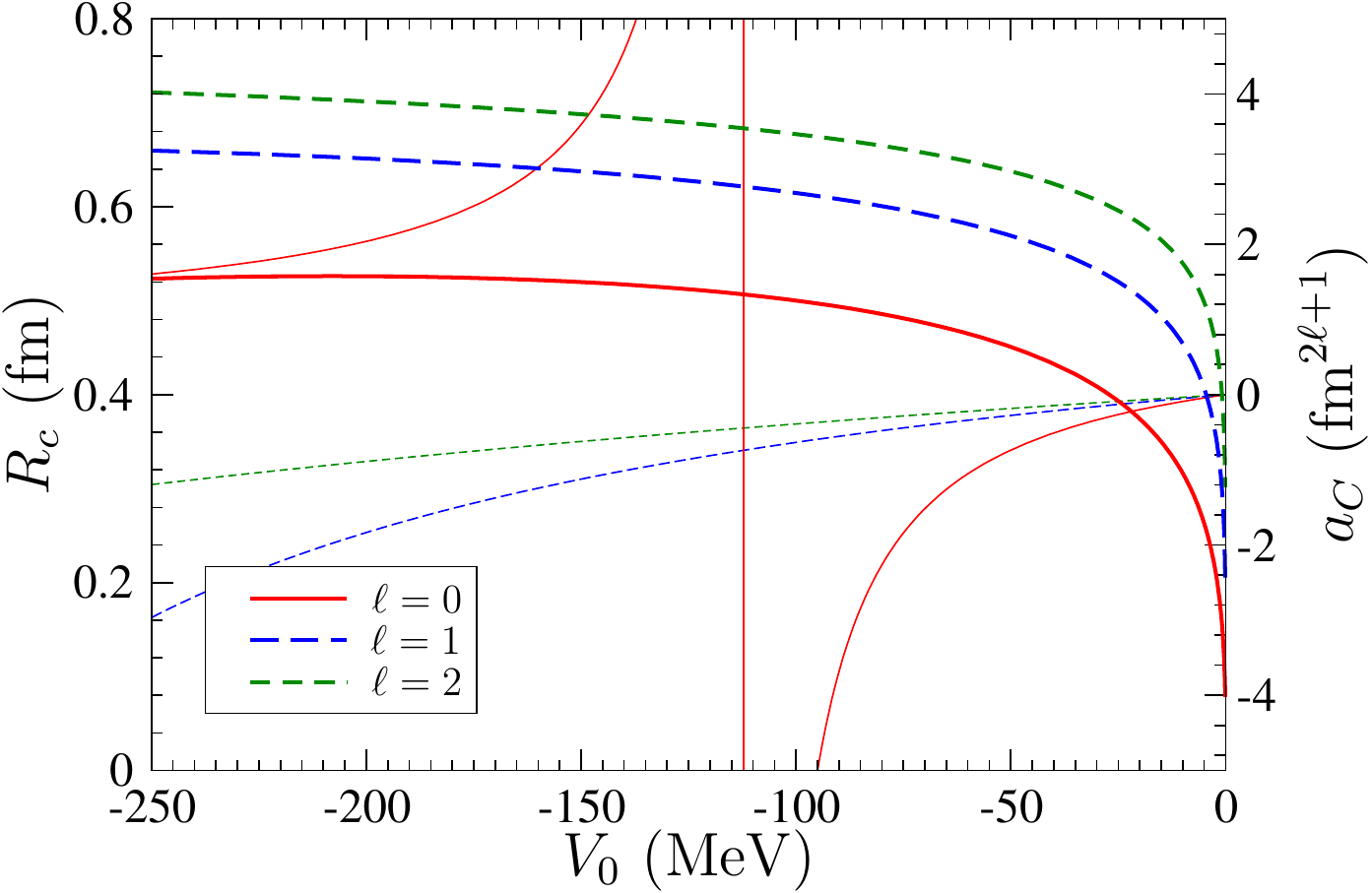}
\caption{Causal range (thick lines) for an attractive step potential and $\gamma
= \gamma_{\text{p--p}}$.  The thin lines show the corresponding scattering
lengths.}
\label{fig:Rc-step-attr-gpp}
\end{figure}

In Figs.~\ref{fig:Rc-step-rep-gpp} and~\ref{fig:Rc-step-attr-gpp} we show the
results (for $\ell=0,1,2$) for both repulsive and attractive step potentials.
Quite interestingly, the $\ell=0$ causal range for the repulsive potential
stays at zero (meaning that one could reproduce the underlying values of the
scattering length and the effective range even with a contact interaction)
until a potential strength of about $100~\MeV$.  For higher partial waves the
causal range takes a nonzero value for much weaker potentials, but the rise is
less steep.  In general, it is remarkable that the causal range is typically
considerably smaller than the actual potential range ($R=1~\fm$).

For attractive potentials the causal range grows much faster as the potential
strength (now negative) increases.  In contrast to what one might expect,
no special features are seen in the causal ranges as the potential becomes
strong enough to support a new bound state close to threshold, \ie, when there
is a pole in the scattering length parameter.

\begin{figure}[htbp]
\centering
\includegraphics[width=0.8\textwidth,clip]{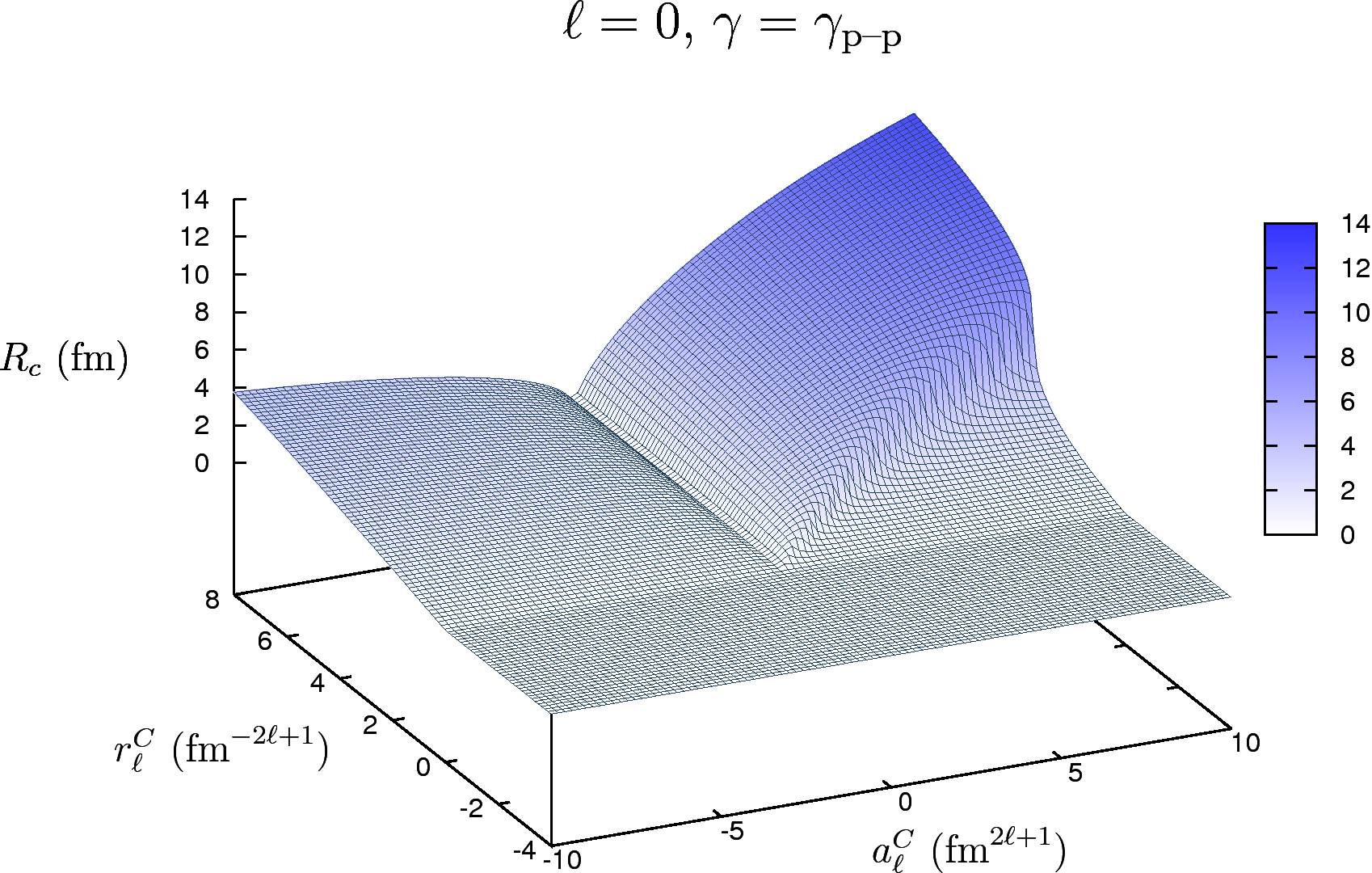}
\caption{Causal range for $\gamma = \gamma_{\text{p--p}}$ and $\ell=0$ in
dependence of $a_0^C$ and $r_0^C$, both measured in $\fm$.}
\label{fig:3dRc-gpp-0}
\end{figure}

\begin{figure}[htbp]
\centering
\includegraphics[width=0.8\textwidth,clip]{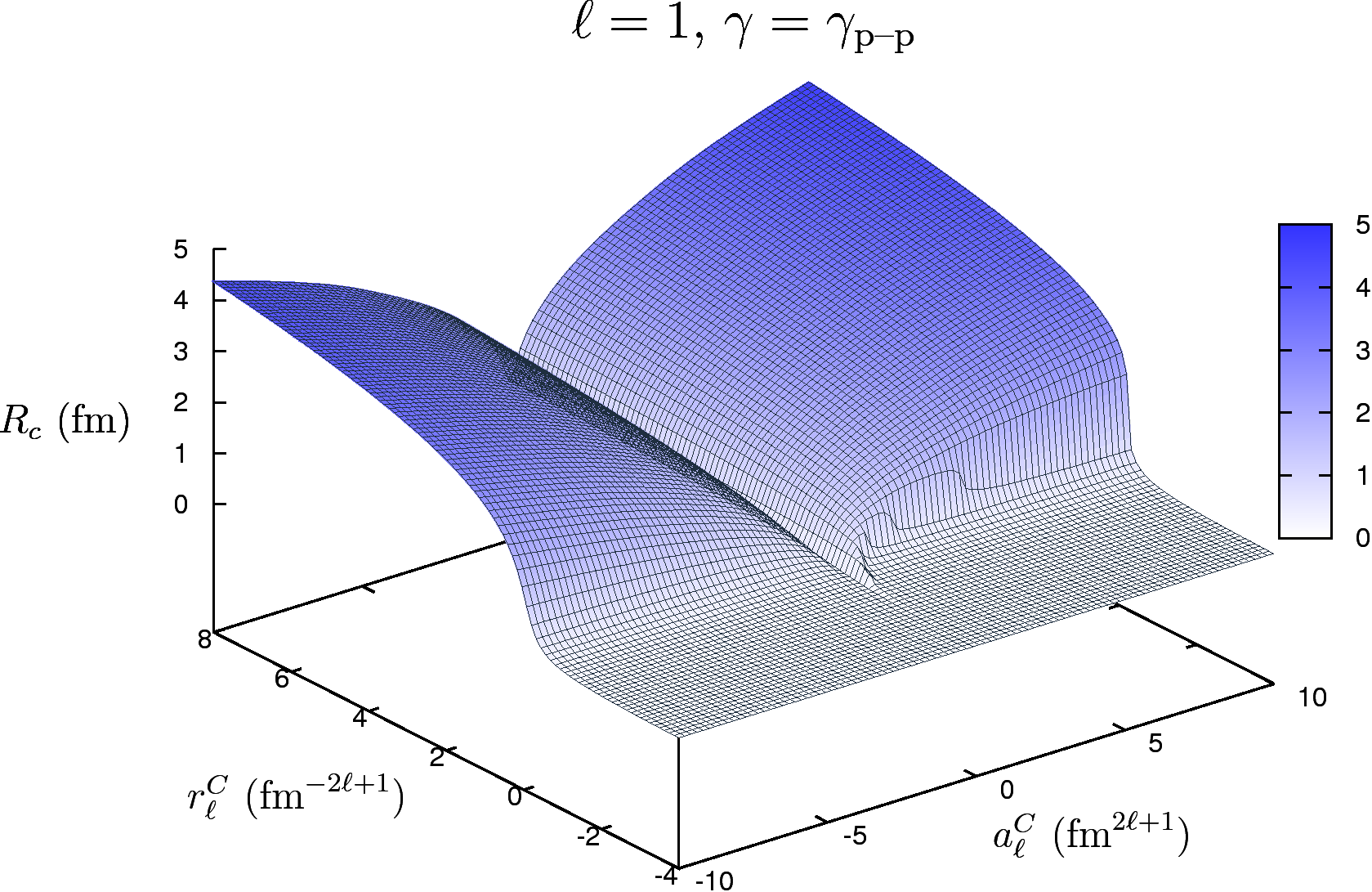}
\caption{Causal range for $\gamma = \gamma_{\text{p--p}}$ and $\ell=1$ in
dependence of $a_1^C$ and $r_1^C$, measured in $\fm^3$ and $\fm^{-1}$,
respectively.}
\label{fig:3dRc-gpp-1}
\end{figure}

\medskip
To conclude this section, we show the general dependence of the causal range on
both the scattering length and the effective range, which has the advantage of
not depending on a certain model potential.  For illustration, we again measure
distances in $\fm$ and set the Coulomb parameter to the value of the
proton--proton system.  In Figs.~\ref{fig:3dRc-gpp-0} and~\ref{fig:3dRc-gpp-1}
we show the results for $\ell=0$ and $\ell=1$.  For negative effective
range $r^{C}$, the causal range stays essentially zero. For positive
effective range, it increases as the absolute value of $a^{C}$ becomes
larger. If one gradually turns off the
Coulomb interaction by letting $\gamma\rightarrow0$, the $\ell=1$ plot stays
almost unchanged, whereas the $\ell=0$ result remains qualitatively the same,
but with a much steeper rise in the quadrant where $a^{C}>0$ and $r^{C}>0$.


\section{Equivalence to relation by Sparenberg~\etal}
\label{sec:Sparenberg-equiv}

In this appendix, we demonstrate the equivalence of our ANC relation to
the one derived by Sparenberg~\etal~\cite{Sparenberg:2009rv}.

Their ANC relation, translated to our notational convention, reads
\begin{equation}
 |A|\frac{\ell!}{\Gamma(\ell+1+\hat{\eta})}
 \approx \kappa^{\ell+1} \sqrt{\tilde{a}_\ell^{C}} \,.
\end{equation}
The tilde on the $\tilde{a}_{\ell}^{C}$ is there to indicate that 
Sparenberg~\textit{et al.} use the convention for the Coulomb-modified effective
range expansion as in
Refs.~\cite{Hamilton:1973xd,Haeringen:1977ab,deMaag:1984ab}, which, as noted at
the end of Sec.~\ref{sec:ERE}, differs from the one used in this paper by an 
overall factor $[2^{\ell}\ell!/(2\ell+1)!]^{2}$.  Combining equations in
Ref.~\cite{Sparenberg:2009rv} and again matching to our notation, one finds that
\begin{equation}
 \frac1{\ell!^2}\prod_{s=1}^\ell(s^2+\eta^2)
 \left[C_{\eta,0}^2\,p^{2\ell+1}\,\cot\tilde{\delta}_\ell(p)
 + \gamma\,p^{2\ell}\,h(p)\right]
 = -\frac1{\tilde{a}^C_\ell}+\frac12\tilde{r}^C_\ell\,p^2 + \cdots \,.
\end{equation}
Note that this is just Eq.~\eqref{eq:ERE-alternative} with the prefactors
combined.  More explicitly, we have
\begin{align}
 \tilde{a}^C_\ell = \left(\frac{\ell!\,2^\ell}{(2\ell+1)!}\right)^2 a^C_\ell
 \mathtext{,} \tilde{r}^C_\ell = \left(\frac{(2\ell+1)!}{\ell!\,2^\ell}\right)^2
 r^C_\ell \mathtext{,\ \ etc.} \,.
\end{align}

\medskip
The final step of the derivation by Sparenberg~\textit{et al.} eliminates the
effective range in favor of the scattering length.  Without invoking this
final step their relation reads
\begin{equation}
 |A|\frac{2^\ell(2\ell+1)!}{\Gamma(\ell+1+\hat{\eta})}
 \approx \kappa^{\ell} \left(-\frac{r^C_\ell}{2}
 + \frac{\gamma^{2\ell-1}}{3(2\ell+1)!^2}\right)^{\!-1/2} \,,
\label{eq:ANC-rel-Sp}
\end{equation}
where we are now using our convention for the effective range expansion.  With
the definition of $\tilde{C}_{\eta,\ell}$ from Eq.~\eqref{eq:C-tilde} and
the values for $\Delta Z_{\ell}$ from Table~\ref{tab:cg2g0} one sees that at
least for $\ell=0,\ldots,2$ this is exactly equivalent to our
Eq.~\eqref{eq:ANC-rel-solved} with the $\OO(\kappa^2)$ set to zero.  In order
to prove the equivalence for arbitrary $\ell$ one would need a general
expression for $\Delta Z_{\ell}$.  This, in turn, requires knowledge of the
constant terms in $W[g_{2},g_{0}](r)$ for arbitrary $\ell$.  It would thus
probably be more interesting to turn the argument around, \ie, take the
equivalence for granted and derive from it a general expression for the constant
terms in the Wronskians.  The only additional ingredient one would need for this
procedure is a general series expansion for the Meijer $G$-functions that arise
from the integral of $g_{0}(r)^{2}$.

\end{document}